\documentclass[aps,nofootinbib,twocolumn,prd,eqsecnum,showpacs,showkeys,preprintnumbers]{revtex4-1}

\usepackage[caption=false]{subfig}
\usepackage{graphicx}
\usepackage{amsmath}
\usepackage{amsfonts}
\usepackage{amssymb}
\usepackage{color}
\usepackage{bm}
\usepackage{mathrsfs}
\usepackage{epstopdf}
\usepackage{url}
\usepackage{footnote}
\usepackage{textcomp}
\usepackage{ulem}

\usepackage{esint}
\usepackage[unicode=true, pdfusetitle,
 bookmarks=true,bookmarksnumbered=false,bookmarksopen=false,
 breaklinks=false,pdfborder={0 0 1},backref=false,colorlinks=false]
 {hyperref}

\usepackage{multirow}

\usepackage{pifont}

\makeatletter
\newcommand*{\rom}[1]{\expandafter\@slowromancap\romannumeral #1@}
\makeatother

\begin{document}

\title{An interacting holographic dark energy model within an induced gravity brane}

\author{Moulay-Hicham Belkacemi$^{1}$}
\email{hicham.belkacemi@gmail.com}
\author{Zahra Bouabdallaoui$^{1}$}
\email{zahraandto@hotmail.com}
\author{Mariam Bouhmadi-L\'{o}pez$^{2,3}$}
\email{mariam.bouhmadi@ehu.eus}
\author{Ahmed Errahmani$^{1}$}
\email{ahmederrahmani1@yahoo.fr}
\author{Taoufik Ouali$^{1}$}
\email{ouali1962@gmail.com}
\date{\today }
\affiliation{
${}^1${\mbox {Laboratory of Physics of Matter and Radiation, Mohammed I University, BP 717, Oujda, Morocco}}\\
${}^2${\mbox {Department of Theoretical Physics, University of the Basque Country UPV/EHU, P.O. Box 644, 48080 Bilbao, Spain}}\\
${}^3${\mbox {IKERBASQUE, Basque Foundation for Science, 48011, Bilbao, Spain}}\\
}


\begin{abstract}
In this paper, we present a model for the late-time
evolution of the universe where a dark energy-dark matter interaction
is invoked. Dark energy is modeled through an holographic Ricci dark
energy component. The model is embedded within an induced gravity brane-world
model. For suitable choices of the interaction  coupling,  the big rip and little rip induced by the holographic Ricci dark
energy, in a relativistic model and in an induced gravity brane-world model, are
removed. In this scenario, the holographic dark energy will have a
phantom-like behaviour even though the brane is asymptotically de
Sitter. 
\end{abstract}

\maketitle
\section{Introduction}

Recent astrophysical observations of type Ia supernovae (SNIa) \cite%
{Perlmutter:1998np}, cosmic microwave background (CMB) \cite{Komatsu:2010fb}, large scale structure (LSS) \cite{Tegmark2004}, and the recent Planck measurements \cite{Planck}, suggest that our
universe is undergoing an era of accelerated expansion. Quantitative
analysis shows that there is a dark energy component (DE) with a negative pressure
component leading to the current accelerating expansion
of the Universe. Since the fundamental origin and nature of such a dark energy  remain
enigmatic at present, various models of dark energy have been put forward, such as a
small positive cosmological constant\ \cite{Peebles} and several kinds of
scalar fields like quintessence \cite{Ratra}, k-essence \cite%
{ARMENDARIZ-PICON2001}, phantom \cite{Caldwell}, etc. 
{ Furthermore, the price of explaining the current cosmic acceleration   by DE is the appearence  of  future singularities at late-time Universe. While some singularities like a big rip (BR) \cite{Caldwell}, sudden singularities \cite{Noj, Bar}, big freeze singularities \cite{Bouh3} and big brake singularities \cite{Chim1,Cata} could happen at a  finite cosmic time, others abrupt which are events smoother like a little rip (LR) \cite{Ruz, Noj2, Stef, Bouh1, Fram, Bre, Bouh2}, a little sibling of a BR \cite{Yas1}, a little bang and a  little sibling of big bang \cite{Yas2} could happen at a infinite cosmic time.}

Currently, another model inspired by the holographic principle has been put
forward to explain the current cosmic acceleration, which states that the number of degrees
of freedom for a system within a finite region should be finite and bounded
by the area of its boundary \cite{'tHooft:1993gx, Susskind:1994vu}.
It is commonly believed that the holographic principle
\cite{'tHooft:1993gx, Susskind:1994vu, Bekenstein} is a fundamental principle
of quantum gravity. Based on an effective quantum field theory, Cohen et
al. \cite{Cohen:1998zx} pointed out that, for a system with size L, which is not a black hole, the quantum vacuum energy of the system
should not exceed the mass of the same size black hole, i.e. $L^{3}\rho
_{\Lambda }\leq LM_{p}^{2}$, where $\rho_{\Lambda }$ is the vacuum energy
density fixed by UV cutoff $\Lambda $ and $M_{p}$ denotes the Plank mass.
The largest IR cutoff $L$ is chosen by saturating the
inequality \cite{Li:2004rb, Hsu:2004ri} so that we get the
holographic energy density $\rho_{H}=3c^{2}M_{p}^{2}/L^{2}$, where $c$
is a numerical factor. Later on and for convenience, we will instead use the parameter defined
as $\beta =c^{2}$. The holographic dark energy model is based on applying the previous ideas to the universe as a whole with the goal of explaining the current speed up of the universe. Then, the IR cutoff can be taken as a
cosmic scale of the universe, like the Hubble horizon \cite{Li:2004rb,Hsu:2004ri},
particle horizon, event horizon \cite{Li:2004rb} or second order geometrical  invariants \cite{luongo1}. Another
choice for the IR cutoff L was suggested by Gao et al. \cite{Gao} (see also \cite{NojiriGRG2006}),
in which the IR cutoff of the holographic dark energy (HDE) is taken
to be the Ricci scalar curvature. Being an invariant quantity, the Ricci scalar curvature has many particularities such as its cosmological implications in describing the HDE \cite{Yas2,Gao,NojiriGRG2006,OualiPRD85}, its space-time dependence and its advantages in avoiding the fine tuning and the causality problems \cite{Li:2004rb}. For all these reasons we will choice the Ricci scalar as the holographic cutoff.

Another approach to explain the observed acceleration of the late universe is
based on alternative theory of general relativity like the one inspired by string theory such as the brane-world scenario. The induced gravity brane-world model proposed by Dvali, Gabadadze,
and Porrati (DGP) is well known and studied \cite{Dvali:2000hr}. It contains two branches \cite{Deffayet:2000uy} the self accelerating branch which suffers from some problems and the normal branch. Even though the normal branch is healthy it cannot describe the current acceleration of the universe unless a dark energy component is invoked \cite{Sahni,Zhang} or the gravitational action is modified \cite{BouhmadiLopez:2010pp}.  In the context of the DGP scenario, different models  have been studied with various kind of sources for DE in Refs \cite{jawad,dutta,Rav,far,she,gha,Shtanov1}. \\

Furthermore, in the context of the dark sector, one of the main problem raised and without explanation in the framework of $\Lambda$CDM cosmology is the cosmic coincidence problem, i.e. why dark energy density is of the same order of magnitude as cold dark matter energy density (CDM). An interacting mechanism between these two components could alleviate the cosmic coincidence problem as suggested by several authors \cite{Bouhmadi-Lopez:2016dcs,Morais:2016bev,AmendolaPRD62, BoehmerPRD78, ChenPRD78, PavonPLB628, CampoPRD78, DuranPRD83, Rav}. These kind of interactions have been invoked in \cite{AbdallaPRD95} in order to explain the possible departure from the $\Lambda$CDM model as measured recently by the experiment of the BOSS \cite{delubacAA574} for a value of the Hubble parameter at redshift $z=2.34$.
In Ref. \cite{OualiPRD85}, we pointed out that the normal branch when filled with an holographic Ricci dark energy (HRDE) can face some DE singularities. This motivated us to improve the model with the aim to remove or smooth these singularities by introducing an interaction between the HDE density and the CDM sector. An interaction between DE and CDM can relieve the coincidence problem and at the same time may smooth some DE singularities. \\
{Recently, interactions between DE and CDM in the holographic model have received a great interest 
by choosing the infrared cutoff as the Hubble scale \cite{PavonPLB628, shey}, as the future event horizon \cite{WangPLB624}, as the Granda and Oliveros scale \cite{jawad}, as a Ricci scale \cite{Yas2} and as a modified holographic DE \cite{Chim2,Chim3}}
We show that this kind of interaction is also a promising way to avoid or to smooth the big rip and the little rip  appearing in the non interacting models \cite{Li:2004rb, Hsu:2004ri, OualiPRD85, BouhmadiPRD84}. \\

The unknown nature of  DE and CDM makes difficult and imprecise the choice of the form of the interaction between them. However, the interaction is usually considered from a phenomenological point of view \cite{WangPLB624,GuoPRD76,OlivaresPRD74,Wang:2016lxa}, from the outset given in \cite{PavonPLB628, CampoPRD78} or from thermodynamical consideration \cite{WangPLB662,PavonGRG41}. The conservation equations have dimensions of energy density divided by unit of time, therefore, the interaction between DE and CDM is expected to lead precisely to this kind of terms on the right hand side (rhs) of their respective continuity equations, i.e. functions of the energy densities of DE and CDM multiplied by a quantity with units inverse of time such as the Hubble scale as have been widely discussed in Refs \cite{WangPLB624, GuoPRD76, WangPLB637, PavonPLB628, CampoPRD74, CampoPRD71, OlivaresPRD74}. The cosmological perturbations in this kind of models have been studied in \cite{HePLB}. It was shown in that work that if the energy density of DE is phantom like, the curvature perturbations are always stable no matter if the coupling is proportional to DE density or to CDM energy density. However, it was shown that if dark energy is of a quintessence nature, the curvature perturbations are unstable unless the coupling of the interaction is proportional to DE density and the range of the coupling takes some specific values.\\

Motivated by the study  of Refs. \cite{HePLB,AbdallaPRD95}, we consider as well this form of interaction i.e. $Q = \lambda_mH\rho_m$, $Q = \lambda_HH\rho_H$, or $Q = H(\lambda_m\rho_m+\lambda_H\rho_H)$ where $Q$ denotes the interaction between the energy densities $\rho_m$ of CDM and
the HRDE component $\rho_H$. The range of the coupling of the interaction, $\lambda_m$ and $\lambda_H$, are determined by observations  \cite{HePRD83,FengPLB665}.  Considering that there is only energy transfer between DE and CDM, the energy transfer is from CDM to DE if $Q<0$ or from DE to CDM if $Q>0$ (see Eqs. (\ref{EOSH}) and (\ref{EOSm})).   \\

The main aim of this paper is to show that an interacting holographic Ricci dark energy (IHRDE) with CDM can describe suitably the late-time acceleration of the universe, and at the same time improves  the model without interaction  by avoiding the big rip and/or little rip happening in the model we studied in Ref. \cite{OualiPRD85}.\par

The outline of this paper is as follows. In Sec. II we briefly present an interacting CDM-HRDE model within a DGP brane-world model.
We assume that the Ricci scalar is the IR cutoff of the holographic energy density. In Sec. III, we study the modified Friedmann equation without a bulk
Gauss-Bonnet (GB) term by analyzing analytically the asymptotic
behavior of the brane and numerically the whole
expansion of the brane. An appropriate choice of the interaction 
coupling $\lambda_{H}$ avoids the big rip and the little rip from the normal
branch and hence the IHRDE gives a satisfactory and an alternative description of
the late time cosmic acceleration of the universe as compared with the HRDE without
the GB term in the bulk in the absence of interaction. Indeed the later
one modifies the big rip and little rip into a big freeze one
while the former removes them definitively.
In Sec. IV, we consider the model
where the bulk contains a GB term.
In this case the asymptotic behaviour of the IHRDE model depends on the sign of a discriminant ${\mathcal{D}}$ which depends on the holographic parameter $\beta $,  the GB term, and the coupling $\gamma$ (see
Eqs. (\ref{factorD})-(\ref{coeffF})). The IHRDE model succeed in removing the big rip and little
rip from the future evolution of the brane. On this case, the brane will evolve
asymptotically as a de Sitter universe for $\gamma =\frac{1}{2\beta_{\lim }}$,
while for $\gamma \neq \frac{1}{2\beta_{\lim }}$, the situation becomes more complicated and depends on
the values of the holographic parameter as well as on the interaction parameter.  At this
regard, as we will show that the coupling between HRDE and CDM plays a crucial role, even more important than the
GB parameter, in removing the
future singularities. Finally, in Sec. V, we conclude.

\section{\ INTERACTING MODEL AND PARAMETER CONSTRAINTS}

We consider a DGP brane-world model, where the bulk contains a GB curvature
term, and the brane contains an induced gravity term on its action \cite{Kofinas:2003rz,BouhmadiLopez:2008nf}.
We restrict our analysis to the normal branch. Assuming only an interaction between CDM and the
holographic component on the brane, the conservation equations of the energy density  read

\begin{eqnarray}
\overset{\cdot }{\rho }_{\mathrm{H}}+3H(1+\omega_{\mathrm{H}})\rho_{%
\mathrm{H}} &=&-Q  \label{EOSH} \\
\overset{\cdot }{\rho }_{\mathrm{m}}+3H\rho_{\mathrm{m}} &=&Q.  \label{EOSm}
\end{eqnarray}
where $H$ \ is the Hubble parameter and $\omega_{\mathrm{H}}$ denotes the equation of state parameter of the holographic
dark energy component.

The modified Friedmann equation in the normal branch of the DGP brane world universe containing an
holographic Ricci dark energy, $\rho_H$, and a CDM component, with energy density $\rho_m$, can be written as
\cite{Kofinas:2003rz,BouhmadiLopez:2008nf,richard}

\begin{equation}
H^{2}=\frac{1}{3M_{p}^{2}}\rho -\frac{1 }{r_{c}}\left( 1+\frac{%
8\alpha }{3}H^{2}\right) H,  \label{friedmann1}
\end{equation}%
where $\rho =\rho_{\mathrm{m}}+\rho_{%
\mathrm{H}}$ is the total cosmic fluid energy density of the brane. The parameter $r_{c}$ is the cross-over scale which determines the
transition from a 4-dimensional (4D) to a 5-dimensional (5D) behaviour, and $\alpha $ is the Gauss-Bonnet parameter.

Furthermore, for a spatially flat Friedmann-Lema\^{\i}tre-Robertson-Walker universe, the Ricci scalar curvature is given by

\begin{equation}
\mathcal{R=}-6\left( \dot{H}+2H^{2}\right),
\end{equation}%
the dot stands for the derivative with respect to the cosmic time of the
brane.\\

As already mentioned in the introduction, $\rho_{\mathrm{H}}$ is related to
the UV cutoff, while $L$ is related to the IR cutoff. Identifying $L^{-2}$
with $-\mathcal{R}/6,$ the energy density of the HRDE is given by \cite{Gao,OualiPRD85}
\begin{equation}
\rho_{\mathrm{H}}=3\beta M_{P}^{2}\left( \frac{1}{2}\frac{dH^{2}}{dx}%
+2H^{2}\right) ,  \label{HRDE}
\end{equation}%
where $x=\ln (a/a_{0})$. The quantities $a$, $a_0$ and $\beta =c^{2}$ are respectively the scale factor, its
present value and  the holographic dimensionless parameter
which, as we will show, plays a significant role in determining the
asymptotic behavior of the HRDE and therefore of the brane evolution.  From now on, the subscript 0 stands for quantities evaluated at the present time.

The modified Friedmann equation\ ({\ref{friedmann1}) }can be further
rewritten as:

\begin{equation}
E^{2}=\Omega_{\mathrm{m}}+\Omega_{\mathrm{H}}-2\sqrt{\Omega_{%
\mathrm{r_c}}}(1+\Omega_{\mathrm{\alpha }}E^{2})E,  \label{friedmann2}
\end{equation}%
where $E(z)=H/H_{0}$, $z$ is the redshift  and
\begin{eqnarray}
\Omega_{\mathrm{m}} &=&\frac{\rho_{\mathrm{m}}}{3M_{p}^{2}H_{0}^{2}}%
,\,\,\,\,\Omega_{\mathrm{r_c}}=\frac{1}{4r_{\mathrm{c}}^{2}H_{0}^{2}},\text{%
\thinspace\ } \\
\,\Omega_{\mathrm{\alpha }} &=&\frac{8}{3}\alpha H_{0}^{2}, \\
\text{\ }\Omega_{\mathrm{H}} &=&\beta (\dfrac{1}{2}\dfrac{dE^{2}}{dx}%
+2E^{2}).  \label{OmegaH}
\end{eqnarray}

 The cosmological parameters of the model are constrained by
evaluating the Friedmann equation ({\ref{friedmann2})} at present

\begin{equation}
1=\Omega_{\mathrm{m}_{0}}+\Omega_{\mathrm{H}_{0}}-2\sqrt{\Omega
_{r_c}}(1+\Omega_{\mathrm{\alpha }}),  \label{present Friedmann}
\end{equation}
i.e. $E(x=0)=1$. By combining this equation and Eq. (\ref{OmegaH}) we obtain:

\begin{equation}
\frac{dE}{dx}\Big|_{x=0}=-2+\frac{\Omega_{\mathrm{H}_{0}}}{\beta }\text{,}%
\label{E'0}
\end{equation}%
On the other hand, given that the universe is currently accelerating the present value of the deceleration parameter, $q=-(1+d\ln E/dx)$, which reads:
\begin{equation}
q_{0}=1-\frac{\Omega_{\mathrm{H}_{0}}}{\beta }.  \label{q0}
\end{equation}%
must be negative, therefore

\begin{equation}
0<\beta =\frac{1-\Omega_{\mathrm{m}_{0}}+2\sqrt{\Omega_{rc}}%
(1+\Omega_{\mathrm{\alpha }})}{1-q_{0}}<\Omega_{\mathrm{H}_{0}}.
\label{beta}
\end{equation}

As we mentioned in the introduction, the following
analysis will be devoted to the normal branch because it requires dark energy and because it is free from the
theoretical problems plugging the self-accelerating branch. By using the constraint ({\ref{present Friedmann})}, it can be shown that 
the holographic parameter $\beta$  verifies 
\begin{equation}
 \frac{%
1-\Omega_{\mathrm{m}_{0}}}{1-q_{0}}<\frac{1-\Omega_{\mathrm{m}_{0}}+2\sqrt{\Omega_{rc}}}{1-q_{0}}<\beta.
\label{inequalities}
\end{equation}%
Taking into account the latest Planck data \cite{Planck}: $\Omega_{\mathrm{m}_{0}}\sim{0.315}$ and $q_{0}\sim -0.558$, therefore the
ratio $\beta_{\textrm{lim}}\doteq\frac{1-\Omega_{\mathrm{m}_{0}}}{1-q_{0}}$ is
of the order $0.44$, which means that the normal branch is characterized by $0.44<\beta$.

 From equations {(\ref{friedmann2}) and (\ref{OmegaH})} the
variation of the dimensionless Hubble rate $E$ with respect to $x$ is

\begin{equation}
\dfrac{dE^2}{dx}=-\frac{2\Omega_{m}+2(2\beta -1)E^2-4\sqrt{\Omega_{r_c}}(1+\Omega_{\alpha
}E^{2})E}{\beta}  \label{variation of E}
\end{equation}

On the other hand, the energy conservation equation {(\ref{EOSm})} gives

\begin{equation}
\dfrac{d\Omega_{m}}{dx}+3\Omega_{m}=\lambda_{H}\Omega_{H}+\lambda
_{m}\Omega_{m}.,  \label{EOSm1}
\end{equation}

Then, using Eq. (\ref{OmegaH}) we obtain

\begin{equation}
\dfrac{d\Omega_{m}}{dx}=\lambda_{H}\left(\dfrac{\beta }{2}\dfrac{dE^{2}}{dx}%
+2\beta E^{2}\right)+\left(\lambda_{m}-3\right)\Omega_{m},  \label{EOSm2}
\end{equation}

and with Eq. (\ref{variation of E}) \ we get

\begin{eqnarray}
\dfrac{d\Omega_{m}}{dx} &=&\Big[2\lambda_{H}\beta -\left( 2\beta
-1\right) (\lambda_{m}-3)\Big] E^{2}  \notag \\
&+&\Big[ \lambda_{H}\beta -\beta (\lambda_{m}-3)\Big] E\dfrac{dE}{dx}
\label{variation of Omega m} \\
&+2&\sqrt{\Omega_{r_c}}(\lambda_{m}-3)(1+\Omega_{\alpha }E^{2})E,
\notag
\end{eqnarray}

The derivative of equation (\ref{variation of E}) gives

\begin{eqnarray}
\dfrac{d^{2}E^{2}}{dx^{2}}&=&-\frac{2}{\beta }\dfrac{d\Omega_{m}}{dx}-\frac{%
2\left( 2\beta -1\right) }{\beta }\dfrac{dE^{2}}{dx}  \notag \\
&+&\frac{4\sqrt{\Omega_{r_c}}}{\beta }(1+3\Omega_{\alpha }E^{2})%
\dfrac{dE}{dx}  \label{Second variation of E}
\end{eqnarray}

Finally, by using Eq. (\ref{variation of Omega m}), we have

\begin{eqnarray}
\dfrac{d^{2}E^{2}}{dx^{2}} &=&\Big[-\lambda_{H}+(\lambda_{m}-3)-2\frac{\left(
2\beta -1\right) }{\beta }\Big]\dfrac{dE^{2}}{dx}  \notag \\
&+2&\Big[-2\lambda_{H}+\frac{\left( 2\beta -1\right) (\lambda_{m}-3)}{\beta }%
\Big]E^{2}  \notag \\
&+&\frac{4\sqrt{\Omega_{r_c}}}{\beta }(1+3\Omega_{\alpha }E^{2})%
\dfrac{dE}{dx}  \label{variation double de E} \\
&-&\frac{4\sqrt{\Omega_{r_c}}(\lambda_{m}-3)}{\beta }(1+\Omega
_{\alpha }E^{2})E  \notag
\end{eqnarray}

In the absence of interaction, i.e. for $\lambda_{H}=\lambda_{m}=0$, the
model coincides with that of Ref. \cite{OualiPRD85}, in which we have
shown that for $\Omega_{\alpha }=0$, the HRDE will have a phantom-like behaviour until it reaches a big rip
singularity for $\beta <1/2$ and a little rip for $\beta =1/2$. In our
present model, in which the interaction between CDM  and HRDE  is included, we can show
numerically that the little rip for $\beta =1/2$, and the
big rip for $\beta_{\lim }<\beta <1/2$ can be avoided which in the previous work \cite{OualiPRD85}
becomes a big freeze by including a GB term in the bulk action. The question that
now arises is for which values of $\lambda_{H}$, $\lambda_{m}$  and $\beta$ these singularities
appearing in that model can be avoided? 
{ To answer this question, we notice that the differential equation ({\ref{variation double de E})} is not linear so it is not easy to solve analytically. However, our numerical analysis of Eq. ({\ref{variation double de E})} shows that its asymptotic behaviour is the same as the one of Eqs. ({\ref{variation of Omega m})} with ${d\Omega_m}/{dx}=0$ i.e. in the far future $\Omega_m$ as well as ${d\Omega_m}/{dx}$ can be neglected. This simplify considerably our task with respect of the above singularities.}
\section{ Model without a Gauss-Bonnet term}

In order to solve equation ({\ref{variation double de E})}, we
consider first the model without a  bulk Gauss-Bonnet term i.e. $\Omega_{\alpha }=0$, therefore
\begin{eqnarray}  \label{variation of E (dgp)}
\dfrac{d^{2}E^{2}}{dx^{2}} &=&\Big[-\lambda_{H}+(\lambda_{m}-3)-2\frac{\left(
2\beta -1\right) }{\beta }\Big]\dfrac{dE^{2}}{dx} \notag\\
&+&2\Big[-2\lambda_{H}+\frac{\left( 2\beta -1\right) (\lambda_{m}-3)}{\beta }%
\Big]E^{2}  \notag \\
&+&\frac{4\sqrt{\Omega_{r_c}}}{\beta }\dfrac{dE}{dx}-\frac{%
4\sqrt{\Omega_{r_c}}(\lambda_{m}-3)}{\beta }E
\end{eqnarray}
and we can write the equation ({\ref{variation of Omega m})} as

\begin{eqnarray}
&&\left[ \lambda_{H}\beta -\beta (\lambda_{m}-3)\right] \dfrac{dE}{dx}
\label{diff eq} \\
&=&-\left[ 2\lambda_{H}\beta -\left( 2\beta -1\right) (\lambda_{m}-3)%
\right] E-2\sqrt{\Omega_{rc}}(\lambda_{m}-3)  \notag
\end{eqnarray}

In the following, we will discus the solution to the above equation for the
model $Q=\lambda_{\mathrm{m}}H\rho_{\mathrm{m}}$, $Q=\lambda_{\mathrm{H}%
}H\rho_{\mathrm{H}}$, and $Q=\lambda_{\mathrm{m}}H\rho_{\mathrm{m}%
}+\lambda_{\mathrm{H}}H\rho_{\mathrm{H}}.$

\subsection{The model $Q=\lambda_{\mathrm{m}}H\rho_{\mathrm{m}}$}

This model corresponds to $\lambda_{H}=0$ and can be split in two cases.
The case $\lambda_{m}=3$ which implies, from Eq. {(\ref{variation of Omega m})}, that $\Omega_{m}$ is always constant. This case is not physically
acceptable as DM behaves as a cosmological constant. And the case $\lambda_{m}\neq 3$ where the holographic parameter plays a crucial
role in determining the asymptotic behaviour of the brane:\\
\subsubsection{\ Asymptotic behavior $\beta =1/2$}
There is a unique solution corresponding to little rip solution supported by the normal branch
\begin{equation}
E=4\sqrt{\Omega_{rc}}x+C_{1},
\end{equation}
where $C_{1}$ is a constant of integration.  \\

\subsubsection{ Asymptotic behavior $\beta \neq 1/2$:} 

Equation (\ref{diff eq}) gives the solution%
\begin{eqnarray}
&&\left\vert \left( 2\beta -1\right) E-2\sqrt{\Omega_{r_c}}%
\right\vert \\
&=&\left\vert \left( 2\beta -1\right) E_{1}-2\sqrt{\Omega_{r_c}}%
\right\vert \exp [-\frac{\left( 2\beta -1\right) }{\beta }(x-x_{1})],  \notag
\end{eqnarray}
where $E_1$ and $x_1$ are integration constants.
\begin{itemize}
\item For $\beta >1/2$, the dimensionless Hubble rate reaches a constant value and the brane is asymptotically de Sitter
\begin{equation}
E_{\infty }=\frac{2}{2\beta -1}\sqrt{\Omega_{rc}}.
\label{Einfinit m}
\end{equation}

We notice that this
asymptotic de Sitter solution is possible only for $\beta \neq 1/2$.
On the other hand, we note that equation ({\ref{EOSm1}) with }$%
\lambda_{H}=0$, gives the solution $\Omega_{m}=\Omega_{\mathrm{m}%
_{0}}e^{(\lambda_{m}-3)x},$ and by substituting it in Eq. ({\ref{friedmann2}%
), }we conclude that with the finite value of the dimensionless Hubble
parameter $E_{\infty }$ of Eq. ({\ref{Einfinit m}),} $\lambda_{m}$ must be
less or equal to $3$ to ensure that$\ \Omega_{\mathrm{H}}$ (and therefore $%
E $) converges asymptotically in the far future to a finite value. So in this
case the energy density of CDM is practically zero at the far future, and the
universe converges asymptotically to a universe filled exclusively with an
HRDE component.

\item For $\beta_{\textrm{lim}}<\beta <1/2$, the dimensionless Hubble rate blows up in the
far future, and it follows a superaccelerated expansion until it hits a big
rip singularity.\\

Therefore, the IHRDE model with $Q=\lambda_{\mathrm{m}}H\rho_{\mathrm{m}}$
has the same asymptotic behaviour as the model analized in Ref. \cite{OualiPRD85}.
So we conclude that  this model does not succeed in removing the big rip and the little
rip singularities happening in the non-interacting model \cite{OualiPRD85}.
\end{itemize}

\subsection{The model $Q=\lambda_{\mathrm{H}}H\rho_{\mathrm{H}}$}
This model corresponds to $\lambda_{{m}}=0$. The holographic $\beta$ parameter determines the
asymptotic behaviour of the brane as follow.\\
\subsubsection{Asymptotic behavior
$\beta =\beta_{\mathrm{LR}}\doteq\frac{3}{2(\lambda_{H}+3)}$} 

In this case Eq. (\ref{diff eq}) gives a little rip solution:

\begin{equation}
E_{\mathrm{LR}}=4 \sqrt{\Omega_{rc}}x+C_{2},  \label{LR solution}
\end{equation}

where $C_{2\text{ }}$is an integration constant.\\

\subsubsection{Asymptotic behavior $\beta \neq \beta_{\mathrm{LR}}$} 

The solution of Eq. (\ref{diff eq}) is given by:

\begin{eqnarray}
&&\left\vert \left[ 2\lambda_{H}\beta +3\left( 2\beta -1\right) \right]
E-6\sqrt{\Omega_{rc}}\right\vert \\
&=&\left\vert \left[ 2\lambda_{H}\beta +3\left( 2\beta -1\right) \right]
E_{1}-6\sqrt{\Omega_{rc}}\right\vert  \notag \\
&&\exp [-\frac{\left[ 2\lambda_{H}\beta +3\left( 2\beta -1\right) \right] }{%
\beta (\lambda_{H}+3)}(x-x_{1})],  \notag
\end{eqnarray}
where $E_1$ and $x_1$ are integration constants.\\

For clarity we divide our analysis in two cases:
\begin{enumerate}
\item { $\beta <\beta_{\mathrm{LR}}$.}\par

The dimensionless Hubble rate blows up in the far future, and it follows a
super accelerated expansion until it hits a big rip singularity.

\item {$\beta >\beta_{\mathrm{LR}}$.}\par

The brane is asymptotically de Sitter, i.e. the Hubble rate reaches a constant value $E_{\infty }$,
\begin{equation}
E_{\infty }=\frac{6 \sqrt{\Omega_{rc}}}{2\beta \lambda_{H}+3\left( 2\beta
-1\right) }.  \label{Einfinit h}
\end{equation}%
 $E_{\infty }$ must be positive. The condition $E_{\infty }>0$, is directly related to the choice of the parameters of the model as can be seen in
Eq. ({\ref{Einfinit h})}. For $\beta =1/2$, the solution reduces to $E_{\infty }=%
6\sqrt{\Omega_{rc}}/\lambda_{H}$ which is finite and no little rip is reached
for a finite $\lambda_{H}$, unless when $\lambda_{H}\longrightarrow 0$, where $E_{\infty }$ blows up and one approaches the model
where there is no interaction \cite{OualiPRD85}, for which the little rip is
inevitable for $\beta =1/2$. For $\beta >1/2$ the asymptotical de Sitter
solution $E_{\infty }$  remains even if $\lambda_{H}=0.$
\end{enumerate}

We conclude that the interacting model acts only on the interval
$\beta_{\textrm{lim}}<\beta \leq 1/2$ of the normal branch, and by choosing $\beta_{%
\mathrm{LR}}=\frac{3}{2(\lambda_{H}+3)}<\beta_{\textrm{lim}}$ one can avoid the
big rip singularity for $\beta_{\textrm{lim}}<\beta <1/2$ and the little rip for $\beta =1/2$
presents in the model without interaction \cite{OualiPRD85}.
We notice that the inclusion of interaction between CDM and
HRDE  gives a satisfactory results as compared with the
inclusion of the GB effect in the HRDE model studied in Ref. \cite{OualiPRD85}.
Indeed, the inclusion of the GB term does not avoid the singularity but it
alters it to a big freeze singularity while the interacting model smooth the singularities by
an appropriate choose of the limiting value of the holographic parameter.

\subsection{The model $Q=\lambda_{\mathrm{H}}H%
\rho_{\mathrm{H}}+\lambda_{\mathrm{m}}H\rho_{%
\mathrm{m}}$}

\subsubsection{$\lambda_{m}=3$}

In this case Eq. (\ref{diff eq}) gives the solution of the dimensionless Hubble rate as
\begin{equation}
 E = E_{1} \exp [-2(x-x_{1})],
\end{equation}
where $E_1$ and $x_1$ are integration constants and it is reduced to a Minkowski one at far future.

\subsubsection{$\lambda_{H}=(\lambda_{m}-3)$ {and} $\lambda_{m}\neq 3$}

In this case the brane has a negative constant dimensionless Hubble rate and it is not physical at late-time

\begin{equation}
E=-2\sqrt{\Omega_{rc}},
\end{equation}

\subsubsection{$\lambda_{H}\neq(\lambda_{m}-3)$ {and} $\lambda_{m}\neq 3$}

\paragraph{Asymptotic behavior $\beta =\beta_{\text{LR}%
}\doteq\frac{3-\lambda_{m}}{2\lambda_{H}-2\lambda_{m}+6%
}$} 

We notice that for $\lambda_{m}\neq 3$ and $\lambda_{H}\neq(\lambda_{m}- 3)$ the
solution leads to a little rip when $\beta =\beta_{\text{LR}}$ and  the Hubble parameter reads

\begin{equation}
E=4\sqrt{\Omega_{rc}}x+C_{3},
\end{equation}

where $C_{3}$ is a constant of integration.\\

\paragraph{ Asymptotic behavior $\beta \neq\beta_{\text{LR}}$} 

The solution of Eq. (\ref{diff eq}) is given by:

\begin{eqnarray}
&&\left\vert \left[ 2\lambda_{H}\beta -\left( 2\beta -1\right) (\lambda
_{m}-3)\right] E+2\sqrt{\Omega_{rc}}(\lambda_{m}-3)\right\vert \notag\\
&=&\left\vert \left[ 2\lambda_{H}\beta -\left( 2\beta -1\right) (\lambda
_{m}-3)\right] E_{1}+2\sqrt{\Omega_{rc}}(\lambda_{m}-3)\right\vert
\notag \\
&&\exp [-\frac{\left[ 2\lambda_{H}\beta -\left( 2\beta -1\right) (\lambda
_{m}-3)\right] }{\lambda_{H}\beta -\beta (\lambda_{m}-3)}(x-x_{1})]
\end{eqnarray}
\begin{enumerate}

\item If {$\beta >\beta_{\text{LR}}$.}

The asymptotic solution corresponds to a de Sitter brane whose Hubble parameter reads
\begin{equation}
E_{\infty }=\frac{-2\sqrt{\Omega_{rc}}(\lambda_{m}-3)}{2\lambda
_{H}\beta -\left( 2\beta -1\right) (\lambda_{m}-3)}.  \label{Einfinit2}
\end{equation}

This asymptotic de Sitter solution gives a finite asymptotic value of $%
\Omega_{\mathrm{H}}$ (see Eq. (\ref{OmegaH})).\\

Notice that equation ({\ref{EOSm1}) can be written }in the limit
where\ $\Omega_{\mathrm{H}}$\ converges asymptotically to a finite
value\ $\Omega_{\mathrm{H}_{\infty }}$ as $(3-\lambda_{\mathrm{m}})\Omega
_{\mathrm{m\infty }}=\lambda_{\mathrm{H}}\Omega_{\mathrm{H}_{\infty }},${\
so }one can conclude that $\lambda_{m}$ must be always less than $3$ in
order to have $\Omega_{\mathrm{m\infty }}>0$.\\

On the other hand, For $\beta =1/2$ the solution (\ref{Einfinit2}) is reduced to
$E_{\infty }=\frac{2\sqrt{\Omega_{rc}}(3-\lambda_{m})}{\lambda_{H}}$ which is
finite and the little rip is avoided for a finite $\lambda_{H}$.
While for $\lambda_{H}\longrightarrow 0$, $E_{\infty }$ blows
up. In this case one approaches the model $Q=\lambda_{\mathrm{m}}H\rho_{%
\mathrm{m}}$ which coincides with the non interaction case \cite{OualiPRD85}.
For $\beta >1/2$, the asymptotic de Sitter solution $E_{\infty }$ is still
present even for $\lambda_{H}=0$ and $\lambda_{m}=0$. For $\beta_{\lim
}<\beta <0.5$, $E_{\infty }$ is positive by choosing $\lambda_{H}>\frac{%
1-2\beta }{2\beta }(3-\lambda_{m})$. Here again by an appropriate choose of
$\lambda_{H}$, one can avoid the big rip singularity.

\item {$\beta <\beta_{\text{LR}}$.}
 The dimensionless Hubble rate blows up in the future and it follows a super accelerating
 expansion until it reaches a big rip singularity.\\
\end{enumerate}

As in the previous subsection, we notice that the interaction between CDM and
an holographic Ricci dark energy density gives
satisfactory results as compared with the inclusion of a Gauss Bonnet term in the bulk \cite{OualiPRD85}.
Indeed, the inclusion of a (GB) term does not avoid the singularity but it
modifies it to a big freeze singularity while the interacting model does.
It is worth noticing, from Eqs. (\ref{diff eq}) and (\ref{Einfinit2}), that the results of the model $Q=\lambda_{%
\mathrm{H}}H\rho_{\mathrm{H}}+\lambda_{\mathrm{m}}H\rho_{\mathrm{m}}$ are
similar to those with $Q=\lambda_{\mathrm{H}}H\rho_{\mathrm{H}}$ by making the
transformation $\lambda_{H}$ $\longrightarrow \frac{3\lambda_{H}}{%
3-\lambda_{m}}$ in the model $Q=\lambda_{\mathrm{H}}H\rho_{\mathrm{H}}$.

\subsection{ Numerical analysis}

The analytical solutions of equation ({\ref{variation of E (dgp)}) are not at all
obvious so a numerical analysis is required.}

\begin{figure}[t]
\begin{center}
\includegraphics[width=0.75\columnwidth]{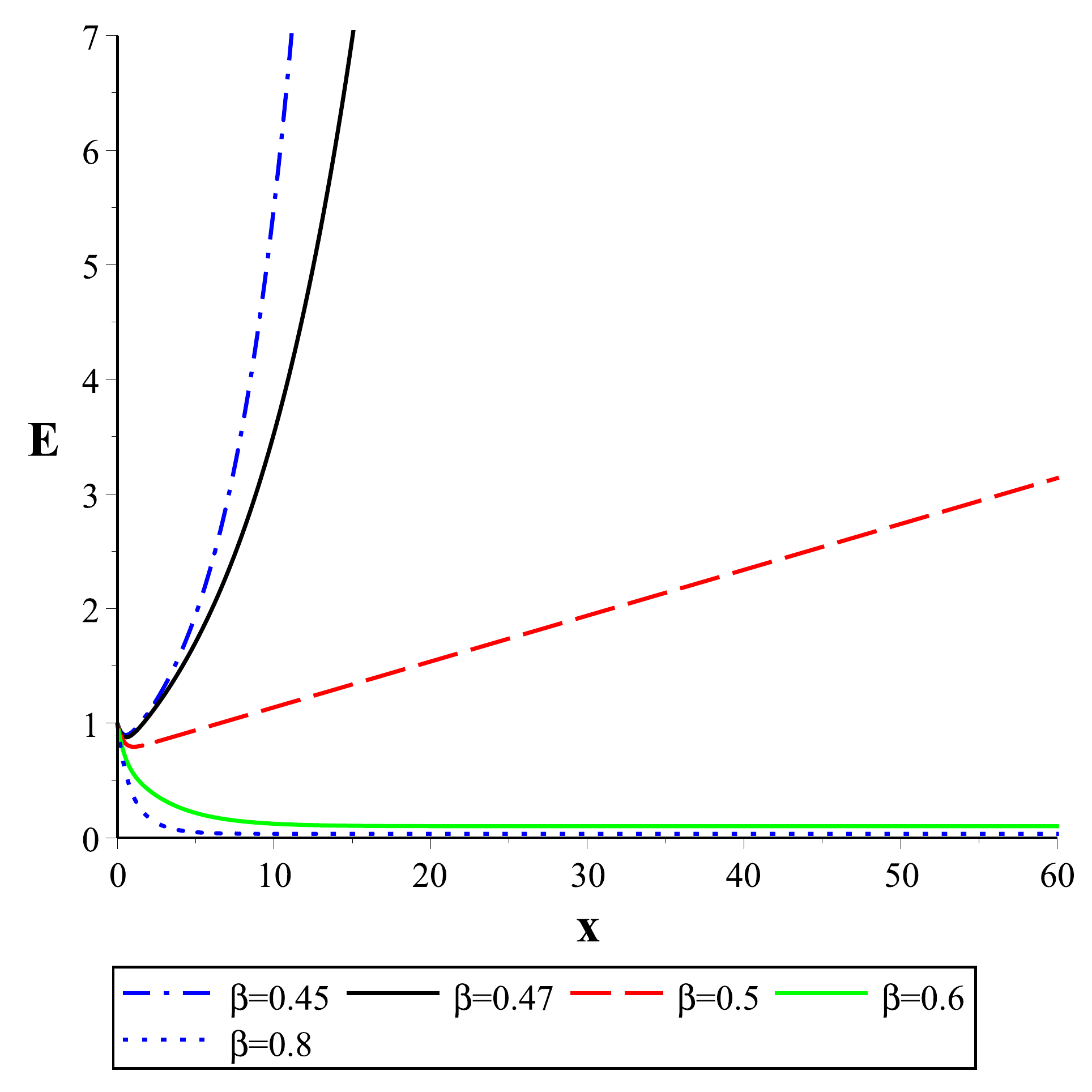}
\end{center}
\caption{{\ Plot of the dimensionless Hubble rate $E$ against $x$. The
cosmological parameters are assumed to be $\Omega_{m_{0}}=0.315$, $%
q_{0}=-0.558 $, $\lambda_{m}=0.1$ and $\lambda_{H}=0$. The
plot has been done for several values of the holographic parameter that are
stated at the bottom of the figure.}}
\label{Em}
\end{figure}
In order to solve numerically Eq. ({\ref{variation of E (dgp)})},
we choose $%
\Omega_{m0}=0.315$ and $q_{0}=-0.558$, as it is given by the latest Planck data
\cite{Planck} and assuming that our model is  pretty much similar to a $%
\Lambda $CDM scenario at the present time. For a given value of $\beta $,
the dimensionless energy density $\Omega_{H_{0}}$ is fixed through Eq. ({%
\ref{q0})}, while the crossover scale parameter $\Omega_{r_c}$ is fixed by
the constraint Eq. ({\ref{present Friedmann})}.

In Fig. \ref{Em}, we show the solutions corresponding to the model $Q=\lambda
_{\mathrm{m}}H\rho_{\mathrm{m}}$ for different values of $\beta $. For $%
{\beta =0.45}$ and $\beta =0.47$ the brane expands exponentially with respect
to $x$ in the future until it reaches a big rip singularity, while for $\beta =0.6$ and $%
\beta =0.8$ the brane expands as an inverse of the exponential of $x$ and remains
asymptotically de Sitter in the future. Finally for $\beta =0.5$
the brane has an asymptotic solution of the form $E=Ax+B$
which corresponds to a little rip solution. All of these numerical
solutions are in agreement with the analytical analysis as it should be.
In addition, the model $Q=\lambda_{\mathrm{m}}H\rho_{\mathrm{m}}$ has the same
asymptotic result as in Ref. \cite{OualiPRD85}.

\begin{figure}[t]
\begin{center}
\includegraphics[width=0.75\columnwidth]{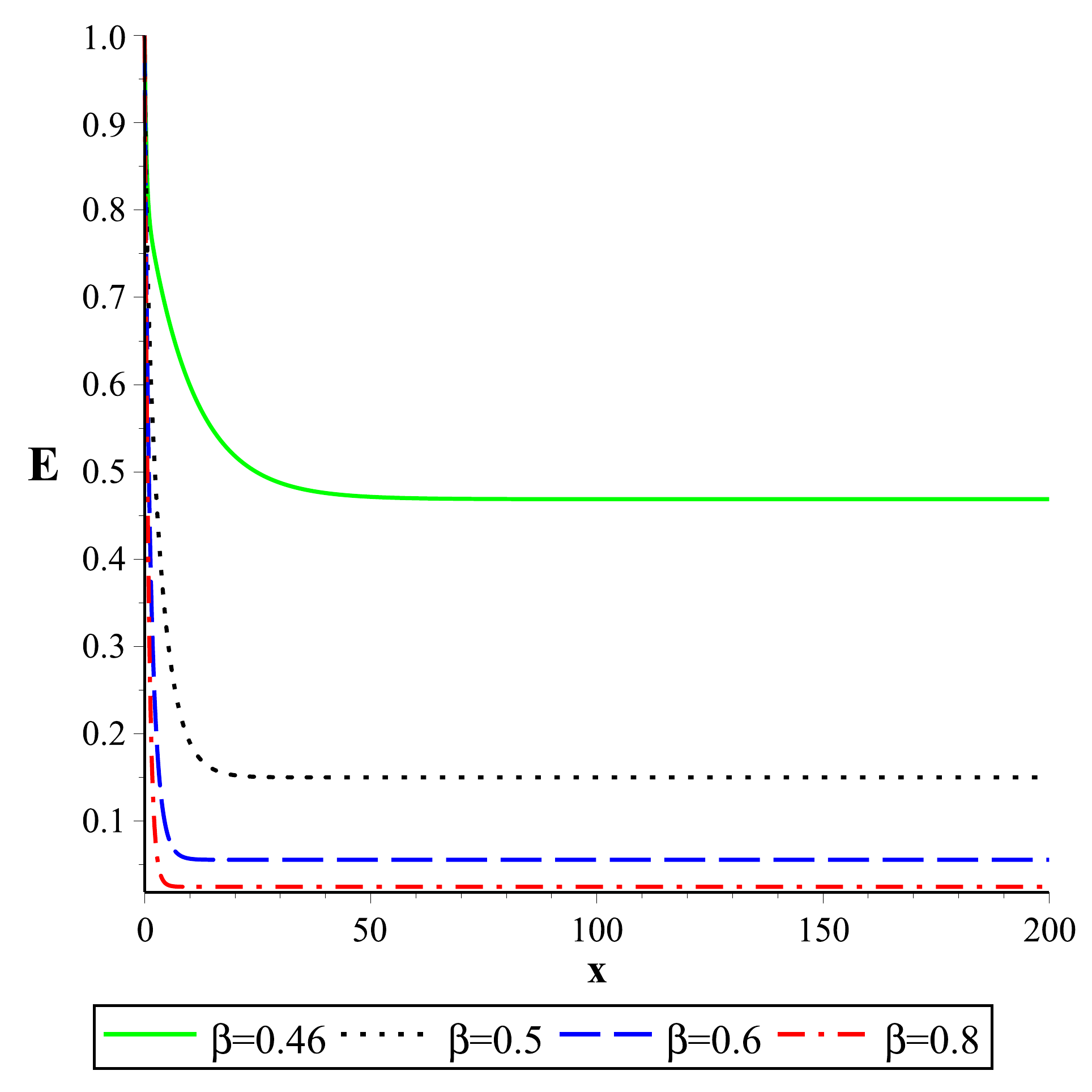}
\end{center}
\caption{{Plot of the dimensionless Hubble rate $E$ against $x$ for the
normal branch. The cosmological parameters are assumed to be $\Omega_{m_{0}}=0.315$, $%
q_{0}=-0.558$, $\lambda_{m}=0$ and $\lambda %
_{H}=0.4$. The plot has been done for several values of the holographic
parameter that are stated at the bottom of the figure.}}
\label{Eh}
\end{figure}

\begin{figure}[t]
\begin{center}
\includegraphics[width=0.75\columnwidth]{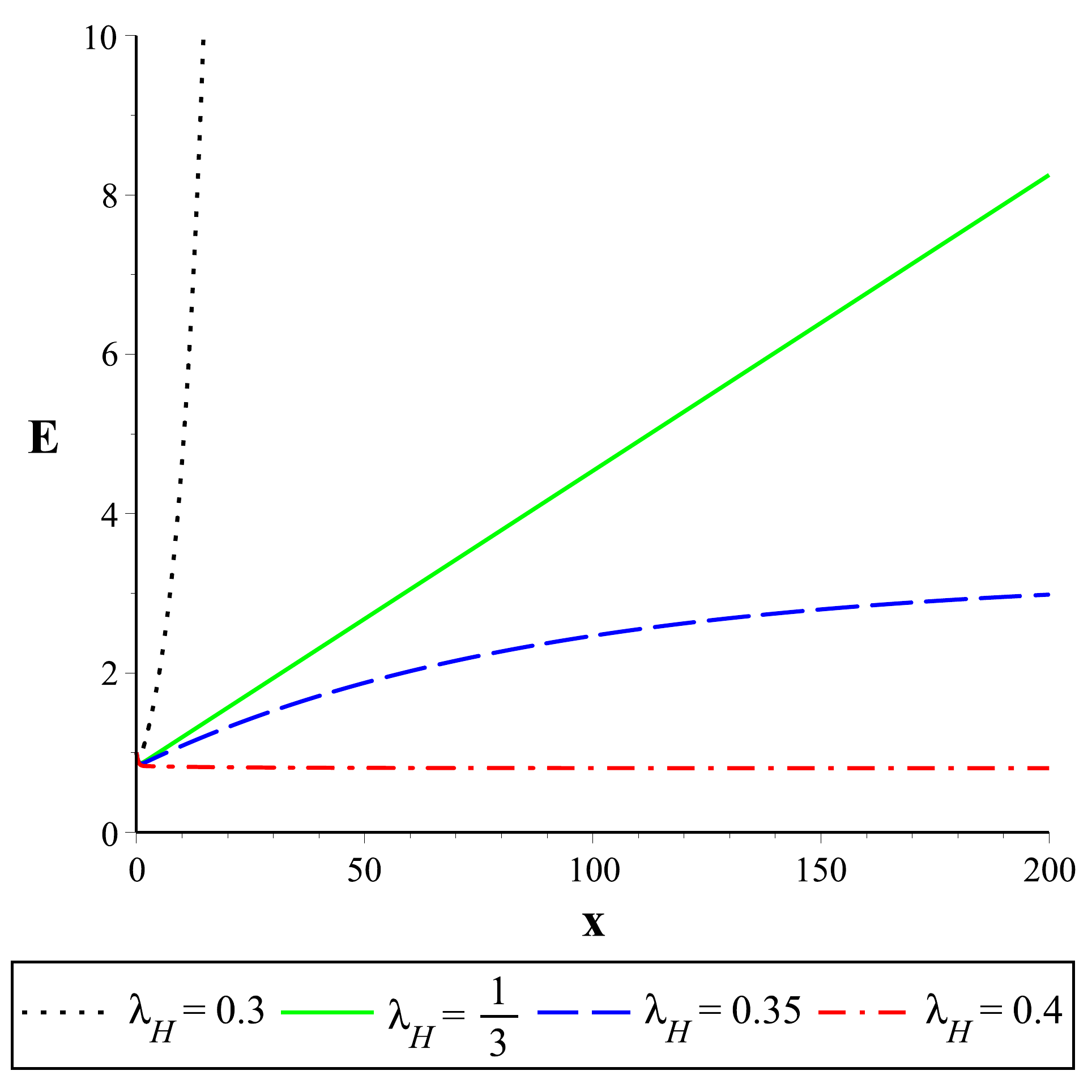}
\end{center}
\caption{{Plot of the dimensionless Hubble rate $E$ against $x$ for the
branch $\beta =0.45$. The cosmological parameters are assumed to be $\Omega_{m_{0}}=0.315$, $%
q_{0}=-0.558 $. The plot has been done for $%
\lambda_{m}=0$ and for several values of the parameter $\lambda %
_{H} $ that are stated at the bottom of the figure. We can see a big rip behaviour for $\lambda_H=0.3$, little rip
behaviour for $\lambda_H=1/3$ and de Sitter behaviour for $\lambda_H\geq 0.35$.}}
\label{E}
\end{figure}

Figs. \ref{Eh} and \ref{E} correspond to the model $Q=\lambda_{\mathrm{H%
}}H\rho_{\mathrm{H}}$. In Fig. \ref{Eh} we show the numerical
solutions of Eq. (\ref{variation of E (dgp)}) for different values of $\beta$.
For all these solutions the condition $\beta >\frac{3}{2(\lambda_{H}+3)}$ is satisfied,\ the brane
is asymptotically de Sitter in the far future and the dimensionless Hubble
rate $E$ approaches the value $E_{\infty }=\frac{6\sqrt{\Omega_{rc}}}{%
2\beta \lambda_{H}+3\left( 2\beta -1\right) }$ which is in agreement with
the analytical analysis (see Eq. (\ref{Einfinit h})).
It is worth noticing the avoidance of the big rip singularity for
$\beta_{\textrm{lim}}<\beta <1/2$ and the little rip singularity for $\beta =1/2$ by
assuming the interaction between CDM and HRDE component, while in the model \cite{OualiPRD85} the big rip
and the little rip singularities are obtained respectively for $\beta <1/2$  and $\beta =1/2$.
Fig. \ref{E} illustrates the numerical solutions of Eq. (\ref{variation
of E (dgp)}) for $\beta =0.45$ and for different values of $%
\lambda_{\mathrm{H}}$. For \ $\lambda_{\mathrm{H}}=1/3$, $\lambda_{\mathrm{H}}>\frac{1}{3}$
and $\lambda_{\mathrm{H}}<\frac{1}{3}$, Fig. \ref{E} shows respectively the little rip solution, the de Sitter behaviour
and the solution   which hits a big rip singularity. For the model $Q=\lambda_HH\rho_H+\lambda_m\rho_m$, the
above discussions are translated to the sets of the couple $(\lambda_m,\lambda_H)$. Indeed, by translating the constraints
 on $\lambda_H$ to $\frac{3\lambda_H}{3-\lambda_m}$ and requiring that the couple $(\lambda_m,\lambda_H)$
 verifies these constraints the same conclusions are obtained.

In order to complete our numerical study {and by imposing that the brane is currently accelerating,} we analyze the equation of state $\omega_{H}=\rm{p}_{H}/\rho_{H}$,  $\omega_{\text{eff}}$
and the deceleration parameter $q$ { where $\rho_H$, $p_H$ and $\omega_{\text{eff}}$ are the HRDE density, its pressure and its effective equation of state associated to the effective energy density (please cf. Eq. (3.16)) respectively. }

From Eq. (\ref{EOSH}) and $Q=\lambda_{\mathrm{H}}H\rho_{\mathrm{H}}$, we
obtain
\begin{equation}
\omega_{H}=-1-\lambda_{H}-\frac{1}{3}\frac{d\ln (\Omega_{H})}{dx},
\end{equation}
where $\Omega_{H}$ is defined in Eq. (\ref{OmegaH}). In terms of the
dimensionless Hubble rate E and its derivatives with respect to $x$, $\omega
_{H}$ can be rewritten as

\begin{equation}
\omega_{\mathrm{H}}=-\lambda_{\mathrm{H}}-1-\dfrac{\left( \dfrac{dE}{dx}%
\right) ^{2}+E\dfrac{d^{2}E}{dx^{2}}+4E\dfrac{dE}{dx}}{3E\dfrac{dE}{dx}%
+6E^{2}}.  \label{eq wH}
\end{equation}
The effective equation of state $\omega_{\text{eff}}$ can be obtained by rewriting the Friedmann equation (\ref{friedmann2}) as the standard Friedmann equation
\begin{equation}\label{eff_friedmann}
    E^2=\frac{1}{3M_p^2H_0^2}(\rho_m+\rho_{\text{eff}}),
\end{equation}
where $\rho_{\text{eff}}$ is the effective energy density that can be defined through Eq. (\ref{HRDE}) as
\begin{equation}\label{}
   \rho_{\text{eff}}= 3M_p^2H_0^2\Big[\frac{1}{2}\beta\frac{dE^2}{dx}+2\beta E^2 -2\sqrt{\Omega_{\mathrm{r_c}}}E\Big].
\end{equation}
Furthermore, from the conservation equation
\begin{equation}
    \dot{\rho}_{\text{eff}}+3H(1+\omega_{\text{eff}})\rho_{\text{eff}}=0, \label{rhoeff1}
\end{equation}
we obtain the effective equation of state
\begin{equation}\label{}
    1+\omega_{\text{eff}}=-\frac{1}{3\rho_{\text{eff}}}\frac{d\rho_{\text{eff}}}{dx}.\label{rhoeff2}
\end{equation}

\begin{figure}[t]
\begin{center}
\includegraphics[width=0.8\columnwidth]{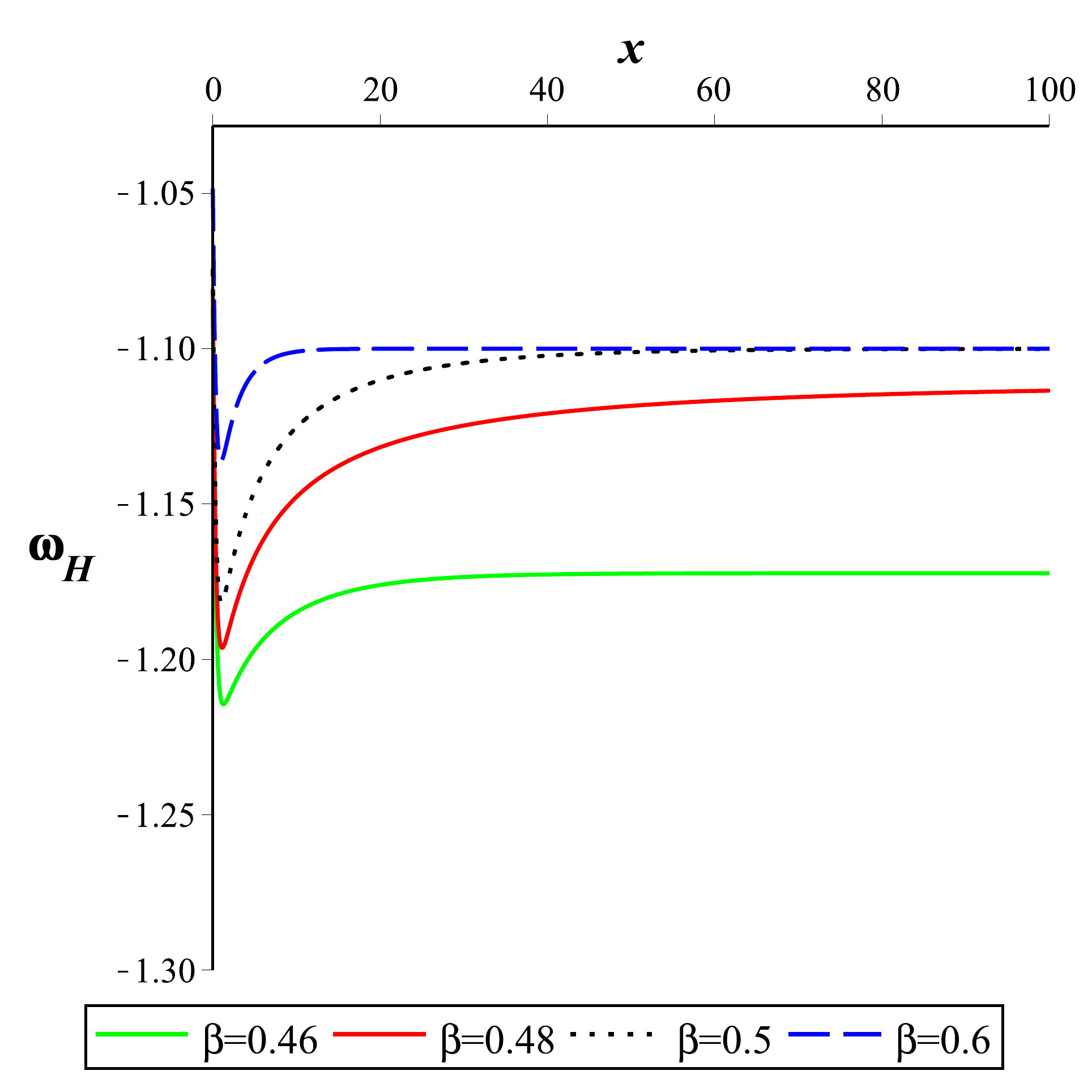}
\end{center}
\caption{{Plot of the equation of state $\omega_H$ against $x$ for $%
\lambda_{H}=0.1$. The cosmological parameters are assumed to be $\Omega_{m_{0}}=0.315$, $%
q_{0}=-0.558$. The plot has been done for several
values of the parameter $\beta$ that are stated at the bottom of the
figure.}}
\label{w}
\end{figure}

\begin{figure}[t]
\begin{center}
\includegraphics[width=0.75\columnwidth]{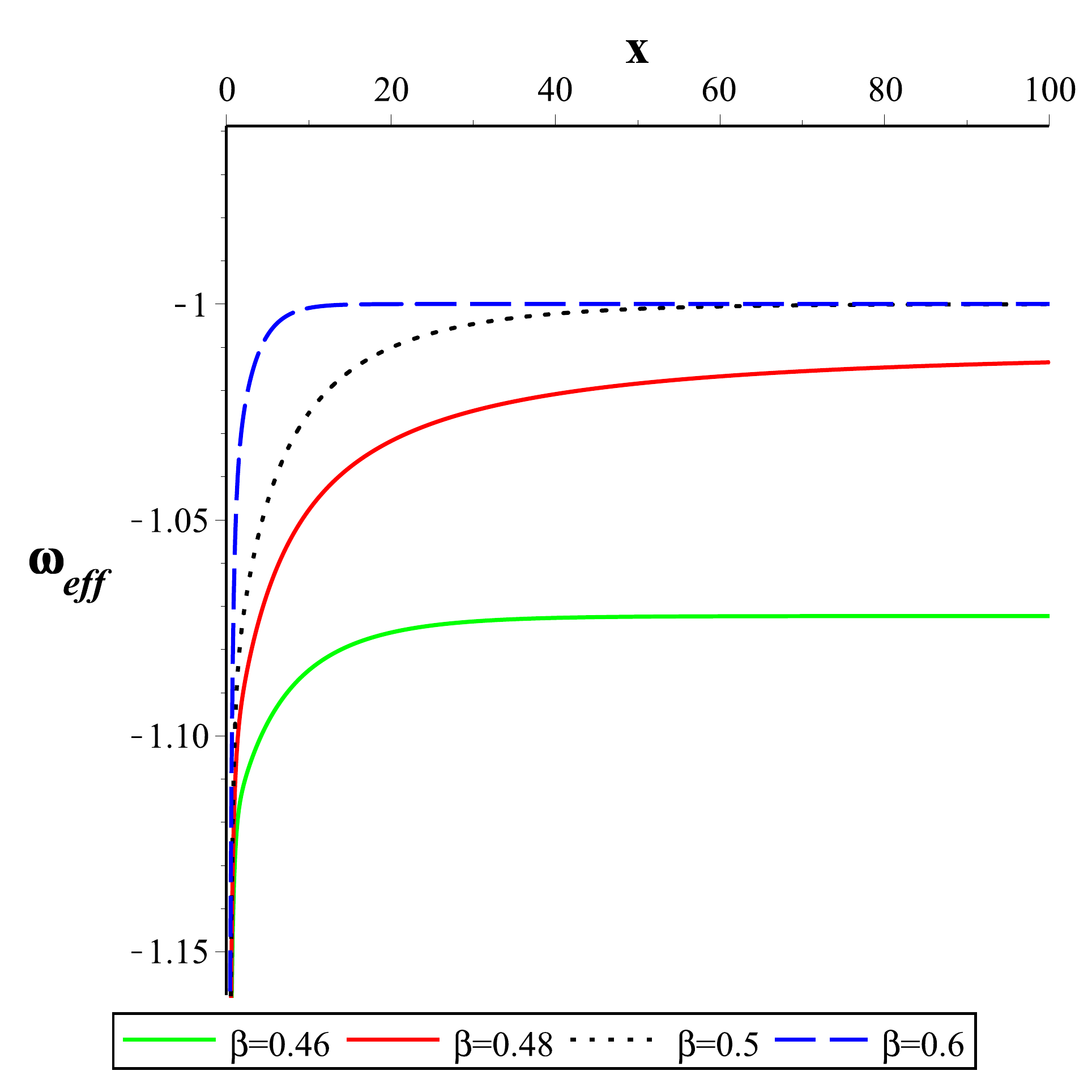}
\end{center}
\caption{{Plot of the effective equation of state $\omega_{\text{eff}}$ against $x$ for $%
\lambda_{H}=0.1$. The cosmological parameters are assumed to be $\Omega_{m_{0}}=0.315$, $%
q_{0}=-0.558$. The plot has been done for several
values of the parameter $\beta$ that are stated at the bottom of the
figure.}}
\label{weff}
\end{figure}

\begin{figure}[t]
\begin{center}
\includegraphics[width=0.75\columnwidth]{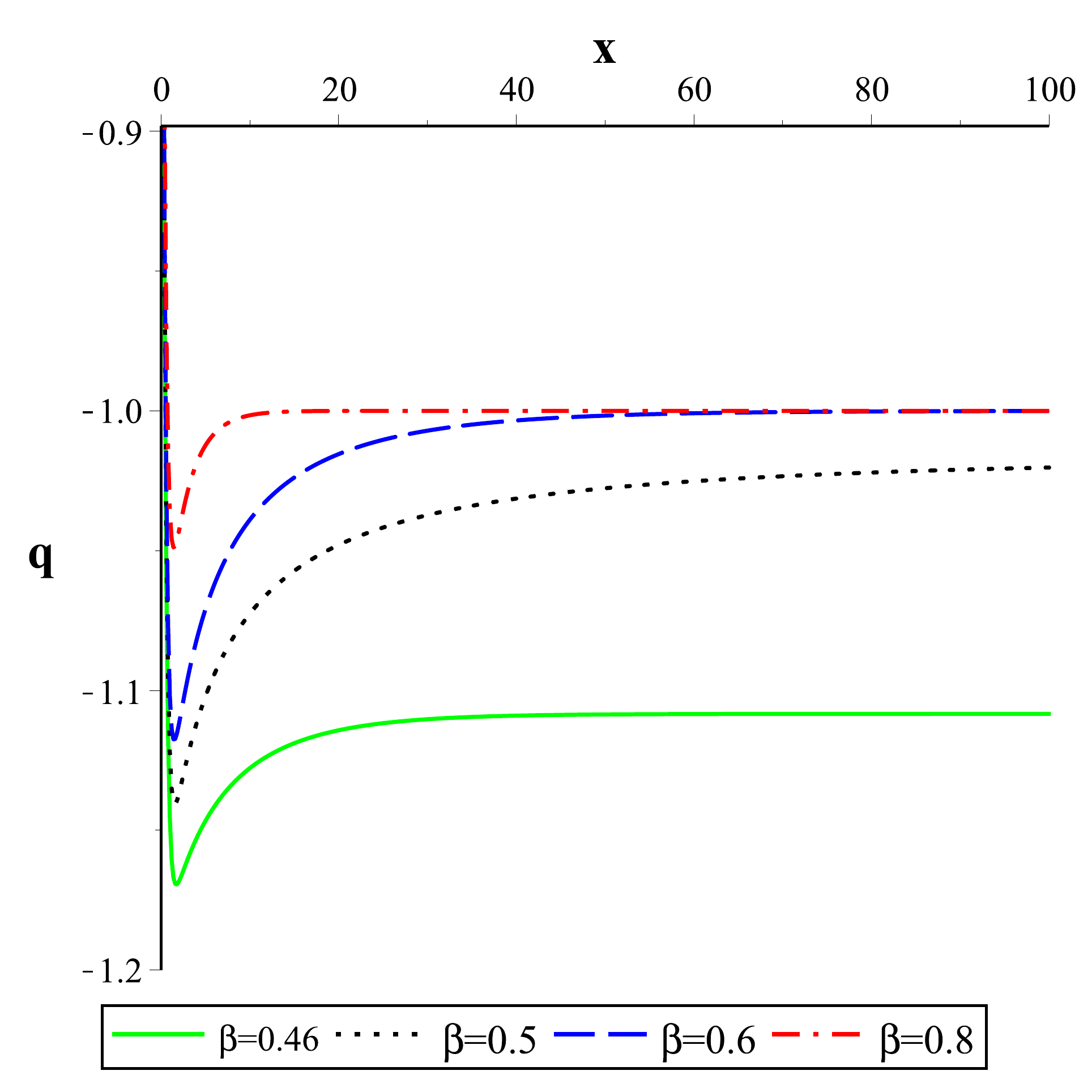}
\end{center}
\caption{{Plot of the deceleration parameter $q$ against $x$ for $%
\lambda_{H}=0.1$. The cosmological parameters are assumed to be $\Omega_{m_{0}}=0.315$, $%
q_{0}=-0.558$. The plot has been done for several values of
the parameter $\beta$ that are stated at the bottom of the figure.}}
\label{Q}
\end{figure}

Fig. \ref{w} shows some examples of the behavior of the equation of
state for the current cosmological values, for $\lambda_{{H}}=0.1$ and for
different values of $\beta $. As it can be clearly noticed from Eq. (\ref{eq
wH}) and illustrated in Fig. \ref{w}, $\omega_{H}$ approaches $-\lambda_{%
\mathrm{H}}-1$ at very late-time for $\beta>\frac{3}{2(\lambda_H+3)}$ corresponding
to a de Sitter behaviour of the brane as it is also confirmed by the effective equation of state $\omega_{\text{eff}}$
plotted in Fig. \ref{weff} 
(for $\lambda_{\mathrm{H}}=0.1$ and $\beta>0.48$, $\omega_{H}$ approaches $-1.1$ and $\omega_{\text{eff}}$ is less than $-1$).
Therefore, HRDE will behave as a phantom
like fluid even though the brane undergoes a de Sitter stage asymptotically. Fig. \ref{Q}
shows that the Universe
continues accelerating in the future. For the model $Q=\lambda_HH\rho_H+\lambda_m\rho_m$, the sets of the couple
$(\lambda_m,\lambda_H)$ that verifie the constraint $\frac{3\lambda_H}{3-\lambda_m}=0.1$ lead to the same conclusions.
We conclude that this kind of interaction can describe the current acceleration expansion of our Universe.

\section{Model with Gauss-Bonnet term in the bulk}

Now we consider the model where the bulk action contains a GB
curvature term in order to analyze the possibility of avoiding the big
freeze present in the absence of interaction between CDM and HRDE and in order
to improve the constraints between the interaction and the beta parameters. In the same manner
we show that the avoidance of the big rip and little rip requires
finite values of $\Omega_{\mathrm{m}}$ at the far future.

From Eq. (\ref{variation of Omega m}), the solution for the interaction $\lambda_{\mathrm{m}}=3$ and $\lambda_{H}=0$
is similar to the case without the GB term which is already analysed in the case $Q=\lambda_{\mathrm{m}}H\rho_{\mathrm{m}}$. Therefore,
Eq. (\ref{variation of Omega m}) should be analyzed only for $\lambda_{\mathrm{m}}\neq 3$ and can be
written as

\begin{eqnarray}
 \Big[ \lambda_{H}\beta -\beta (\lambda_{m}-3)\Big] \dfrac{dE}{dx}
&=&- \Big[ 2\lambda_{H}\beta -\left( 2\beta -1\right) (\lambda_{m}-3)%
\Big] E \nonumber\\
-&2&\sqrt{\Omega_{rc}}(\lambda_{m}-3)(1+\Omega_{\alpha }E^{2})
\label{diff eq with GB}
\end{eqnarray}
in the asymptotic regime where $d\Omega_m/dx\rightarrow 0$.\par
For $\lambda_{\mathrm{m}}\neq 3$, and in order to compare our results to
the case without interaction \cite{OualiPRD85}, it is interesting to rewrite
equation ({\ref{diff eq with GB}) as}

{%
\begin{equation}
\tilde{\beta} \dfrac{dE}{dx}=-\left[ {\left( 2\tilde{\beta}-1\right)
E-2\sqrt{\Omega_{rc}}(1+\Omega_{\alpha }E^{2})}\right] ,
\label{FriedasympGB}
\end{equation}}

where $\tilde{\beta} =\gamma \beta ,$ and
\begin{equation}
\ \gamma =\left(1+\frac{\lambda_{H}}{3-\lambda_{m}}\right).  \label{gamma}
\end{equation}

\begin{figure}[t]
\begin{center}
\includegraphics[width=0.75\columnwidth]{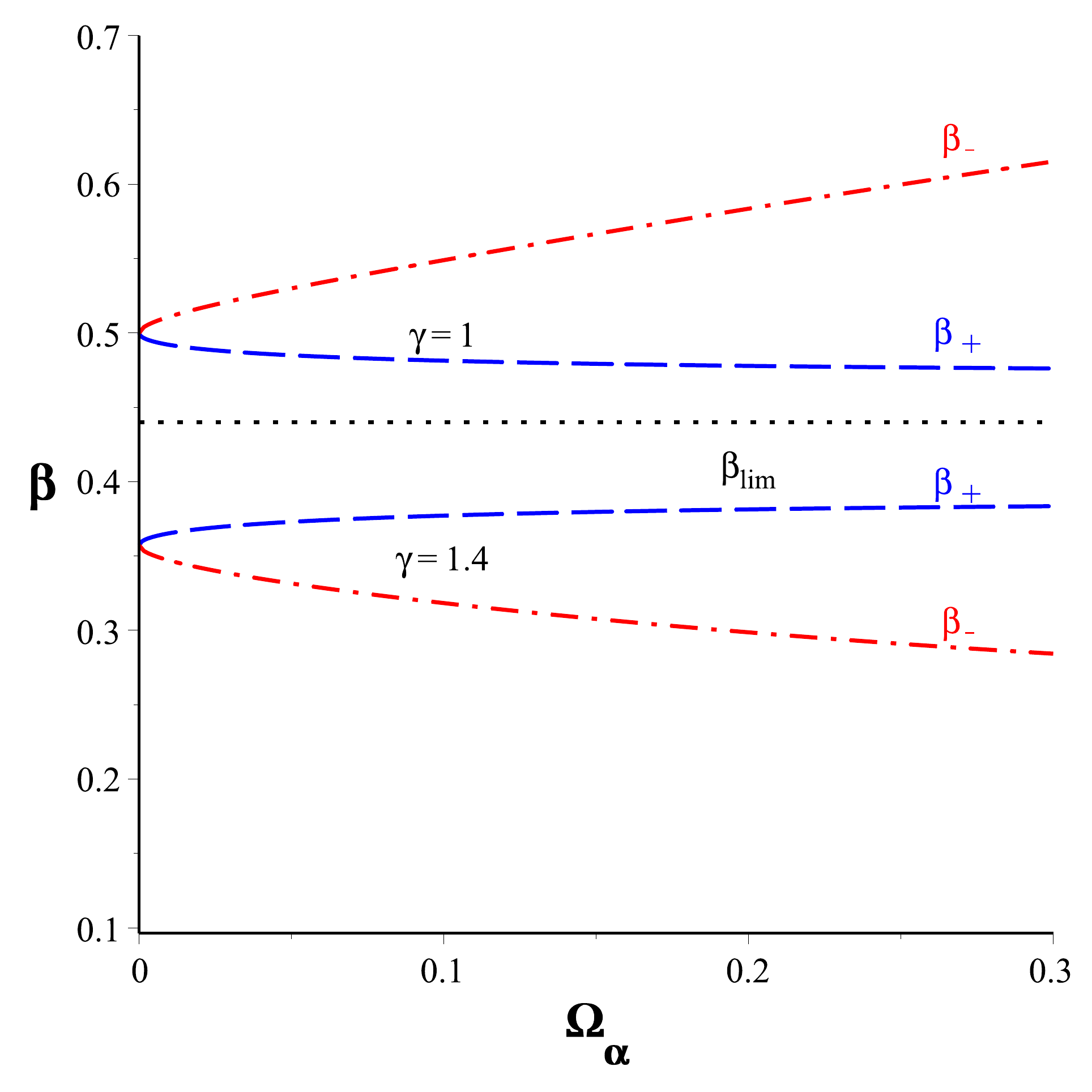}
\end{center}
\caption{{Plot of the $\beta_\pm$ parameters against $\Omega_%
\alpha$. The cosmological parameters are assumed to be $\Omega_{m_{0}}=0.315$, $%
q_{0}=-0.558$. The plot has been done for several values of the parameter $%
\gamma$.}}
\label{beta_pm}
\end{figure}

The solutions of Eq. (\ref{FriedasympGB}) depends  on the sign of the
discriminant $\mathcal{D}$ of the polynomial on its rhs, which
reads
\begin{equation}
\mathcal{D}=\left( 2\tilde{\beta}-1\right) ^{2}-16\Omega_{r_{c}}\Omega
_{\alpha },  \label{D}
\end{equation}%
  and can be factorised as follows
\begin{equation}
\mathcal{D}=\mathcal{F}(\beta  -\beta_{-})(\beta -\beta_{+}),
\label{factorD}
\end{equation}

where

\begin{equation}
\beta_{\pm }=\frac{1+\Omega_{\alpha }\pm 2\sqrt{\Omega_{\alpha }}%
(1-\Omega_{m_0})}{2[(1+\Omega_{\alpha })\gamma \pm \sqrt{\Omega_{\alpha }}%
(1-q_{0})]}.\label{betapm}
\end{equation}

and

\begin{equation}
\mathcal{F}=4\left[\gamma ^{2}-\Omega_{\alpha }\left(\frac{1-q_{0}}{1+\Omega_{\alpha
}}\right)^{2}\right].  \label{coeffF}
\end{equation}
Fig. \ref{beta_pm} shows the behaviour of the parameters $\beta_\pm$ against $\Omega_\alpha$
for the cosmological parameters $\Omega_{m0}=0.315$ and $q_0=-0.558$.
As can be seen from Fig. \ref{beta_pm}, $\beta_+=\beta_-=\beta_{\textrm{lim}}$ for $\gamma=\frac{1}{2\beta_{\textrm{lim}}}$.
For $\frac{1}{2\beta_{\textrm{lim}}}<\gamma$ ($\gamma<\frac{1}{2\beta_{\textrm{lim}}}$), $\beta_+<\beta_-<\beta_{\textrm{lim}}$
($\beta_{\textrm{lim}}<\beta_+<\beta_-$). \par
In the following we analyze the effect of the interaction between CDM and HRDE component on the
singularities appearing on the same model without such interaction (cf. Ref. \cite{OualiPRD85}).

\subsection{$Q=\lambda_{\mathrm{m}}H\rho_{\mathrm{m}}$}

In this case and from Eq. (\ref{gamma}) $\gamma =1$, we obtain the
same asymptotic result as in Ref. \cite{OualiPRD85}, and we conclude, as in
the previous section, that the IHRDE model with $Q=\lambda_{\mathrm{m%
}}H\rho_{\mathrm{m}}$ does not succeed to remove the big freeze presented in
the range $\beta_{\lim }<\beta <\beta_{-}$ for the holographic Ricci dark
energy in a DGP brane world model with a GB term in the bulk.

\begin{figure}[t]
 \begin{center}
 \includegraphics[width=0.75\columnwidth]{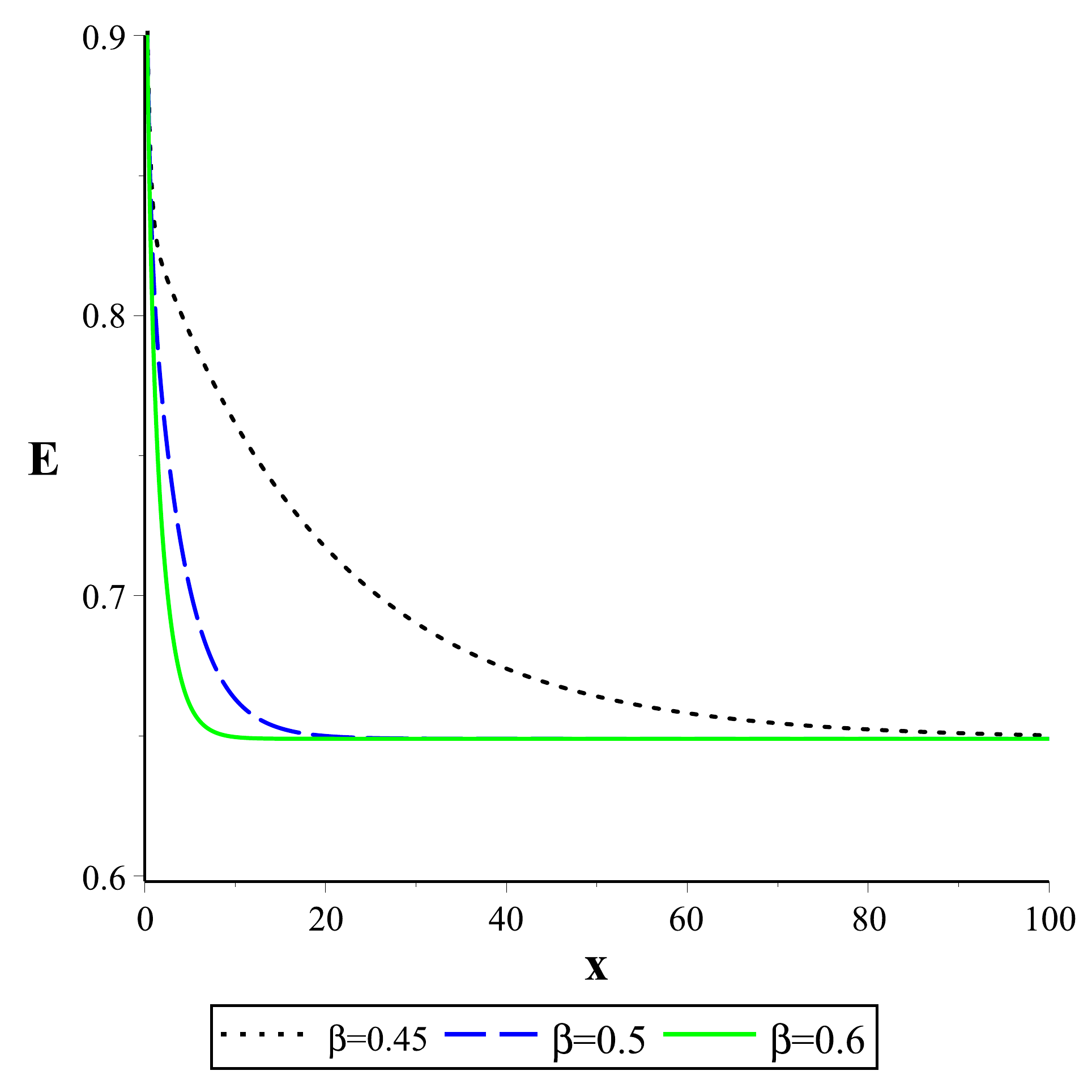}
 \end{center}
 \caption{{Plot of the dimensionless Hubble rate $E$ against $x$
 for an interacting coefficient $\gamma =\frac{1}{2%
 \beta_{\lim }}$. The cosmological parameters are assumed to be $\Omega_{m_{0}}=0.315$, $%
q_{0}=-0.558$ and $\Omega_{\alpha }=0.1$.
 The plot has been done for several values of the holographic parameter that
 are stated at the bottom of the figure.}}
 \label{gamma0}
 \end{figure}
\begin{figure}[t]
\begin{center}
\includegraphics[width=0.75\columnwidth]{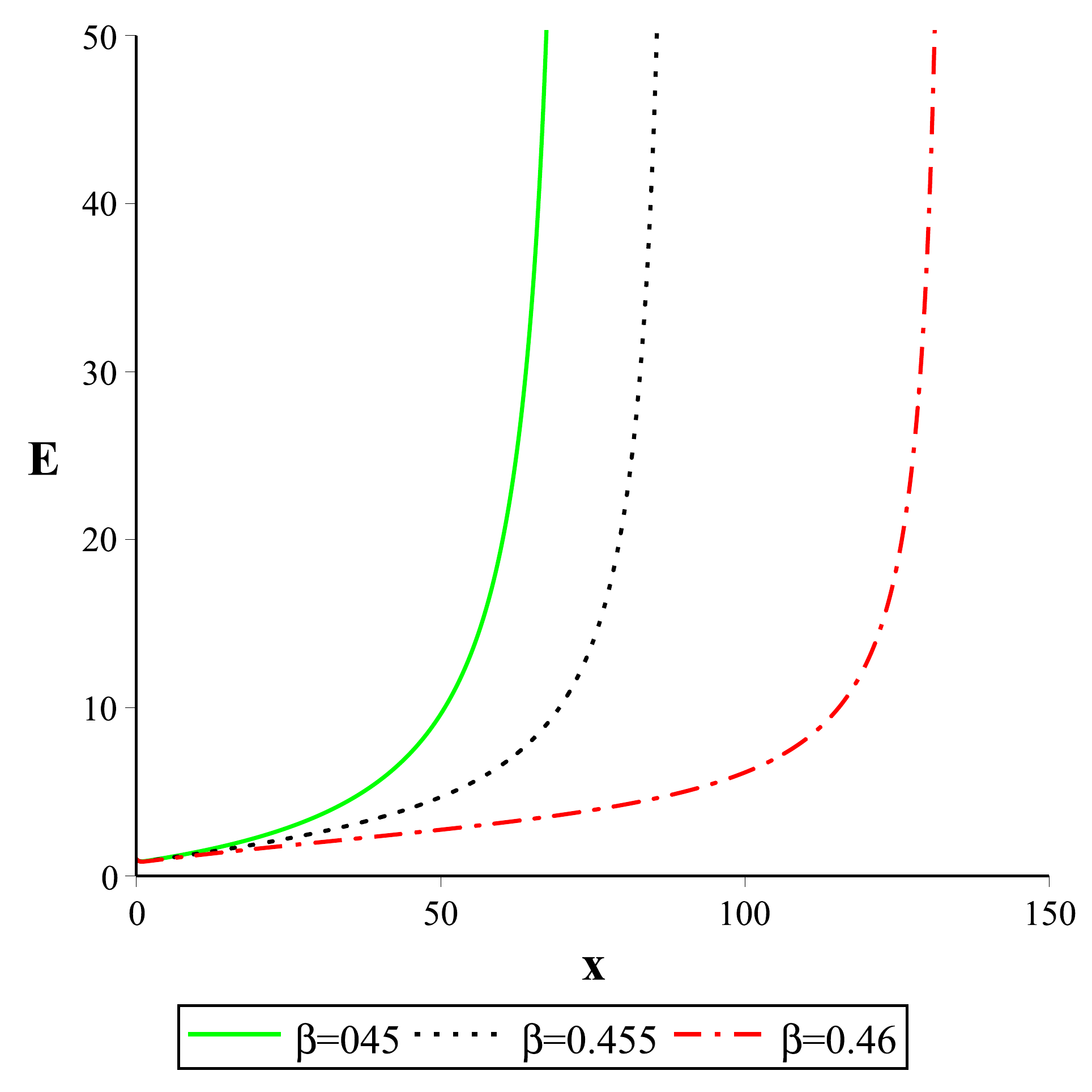}
\end{center}
\caption{Plot of the dimensionless Hubble rate $E$ against $x$ for the
normal branch, and for the negative sign of $\mathcal{D}$. The cosmological parameters
are assumed to be $\Omega_{m_{0}}=0.315$, $%
q_{0}=-0.558$, $\Omega_{%
\alpha }=0.1$, $\lambda_{m}=0$ and $\lambda_{H}=0.3$. The
plot has been done for several values of the holographic parameter that are
stated at the bottom of the figure.}
\label{Dn}
\end{figure}
\subsection{$Q=\lambda_{\mathrm{H}}H\rho_{\mathrm{H}}$ and
$Q=\lambda_{\mathrm{H}}H\rho_{\mathrm{H}}+\lambda %
_{\mathrm{m}}H\rho_{\mathrm{m}}$}
The analysis will be done for the model $Q=\lambda_{\mathrm{H}}H\rho_{\mathrm{H}}$ while
the generalization to the model $Q=\lambda_{\mathrm{H}}H\rho_{\mathrm{H}}+\lambda_{\mathrm{m}}H\rho_{\mathrm{m}}$,
as it can be noticed from Eq. (\ref{diff eq}),
will be obtained  by replacing $\lambda_H$ by $\frac{3\lambda_H}{3-\lambda_m}$
\subsubsection{$\gamma =\frac{1}{2\beta_{\lim }}$}

In this case $\beta_{\lim }=\beta_-=\beta_+$ and the discriminant is equal to $\mathcal{D=}$ $\mathcal{F}(\beta
-\beta_{\lim })^{2}$ which is always positive. The solution of the
asymptotic Friedmann equation (\ref{FriedasympGB}) reads:
\begin{equation}
\left\vert \frac{4\Omega_{\alpha }\sqrt{\Omega_{r_{c}}}E
-2\gamma \beta +1+\sqrt{\mathcal{D}}}{4\Omega_{\alpha }\sqrt{%
\Omega_{r_{c}}}E-2\gamma \beta +1-\sqrt{\mathcal{D}}}\right\vert =%
\mathcal{C}_{2}\exp \left[ -\frac{\sqrt{\mathcal{D}}}{\gamma \beta }(x-x_{2})%
\right] ,
\end{equation}%
where
\begin{equation}
\mathcal{C}_{2}=\left\vert \frac{4\Omega_{\alpha }\sqrt{\Omega
_{r_{c}}}E_{2}-2\gamma \beta +1+\sqrt{\mathcal{D}}}{4\Omega
_{\alpha }\sqrt{\Omega_{r_{c}}}E_{2}-2\gamma \beta +1-\sqrt{\mathcal{D}}}%
\right\vert ,
\end{equation}%
$x_{2}$ and $E_{2}$ are integration constants. We can deduce that, for
very large values of $x$, the brane behaves asymptotically like an expanding de Sitter
universe, i.e. with a  constant positive Hubble rate (see Eq. (\ref{D}))
\begin{equation}
E_{+}=\frac{(2\gamma \beta -1)-\sqrt{\mathcal{D}}}{4\Omega_{\alpha }\sqrt{%
\Omega_{r_{c}}}}=\frac{(\frac{\beta }{\beta_{\lim }}-1)-\sqrt{\mathcal{D}}%
}{4\Omega_{\alpha }\sqrt{\Omega_{r_{c}}}}.
\end{equation}

\subsubsection{$\gamma \neq \frac{1}{2\beta_{\lim }}$}
\begin{enumerate}
	\item {Negative discriminant ($\mathcal{D}<0$):} 
The discriminant $\mathcal{D}$ is negative for $\frac{1}{2\beta_{\lim}}<\gamma$
where $\beta $ satisfies $\beta_{-}<\beta <\beta_{+}<\beta_{\lim }$ or for $\gamma <\frac{1}{2\beta_{\lim }}$
where $\beta_{\lim }<\beta_{+}<\beta <\beta_{-}$  (cf. Eq. (\ref{factorD}) and Fig. \ref{beta_pm}). The
first case will be ignored since it corresponds to the self-accelerating
branch. Hence only the case $\gamma <\frac{1}{2\beta_{\lim }}$ will be
analysed. The Friedmann Eq. (\ref{FriedasympGB}) can be integrated as

\begin{equation}
E=\frac{1}{4\Omega_{\alpha }\sqrt{\Omega_{rc}}}\left\{ \sqrt{\left\vert
\mathcal{D}\right\vert }\tan \left[ \frac{\sqrt{\left\vert \mathcal{D}%
\right\vert }}{2\gamma \beta }\left( x-\mathcal{C}_{2}\right) \right]
+2\gamma \beta -1\right\} ,  \label{solution D<0}
\end{equation}%
where
\begin{equation}
\mathcal{C}_{2}=x_{2}+\frac{2\gamma \beta }{\sqrt{\left\vert \mathcal{D}%
\right\vert }}\arctan \left[ \frac{-4\Omega_{\alpha }\sqrt{\Omega_{r_{c}}}%
E_{2}+2\gamma \beta -1}{\sqrt{\left\vert \mathcal{D}\right\vert }}\right] ,
\end{equation}%
$x_{2}$ and $E_{2}$ are integration constants. Consequently, there is
always a finite value of the scale factor or $x$ {where the Hubble rate and its derivative blow
up at}
\begin{equation}
x_{\mathrm{sing_{1}}}=\mathcal{C}_{2}+\frac{2\gamma \beta }{\sqrt{\left\vert
\mathcal{D}\right\vert }}\left(n+\frac{1}{2}\right)\pi ,\quad n\in \mathbb{Z},
\end{equation}%
where $"{\mathrm{sing_{1}}}"$ denotes the big freeze singularity.
Therefore we conclude that the brane hits a big freeze singularity in the
future as the event $x_{\mathrm{sing_{1}}}$ takes place at a finite future
cosmic time $t_{\mathrm{sing_{1}}}$ \cite{Noj,BouhmadiLopez:PLB2008}.\\
\begin{figure}[t]
\begin{center}
\includegraphics[width=0.75\columnwidth]{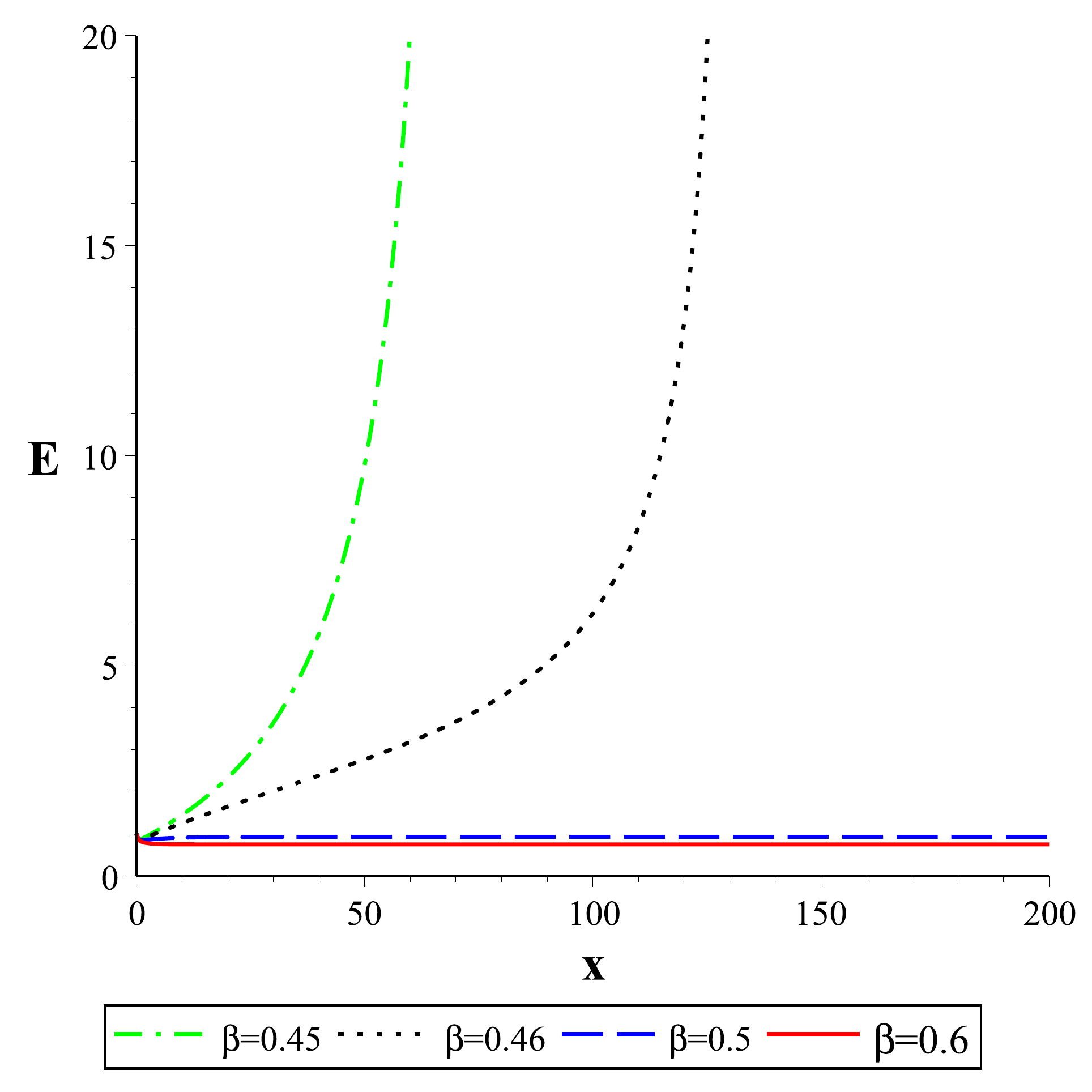}
\end{center}
\caption{Plot of the dimensionless Hubble rate $E$ against $x$, for the
positive sign of $\mathcal{D}$, and for an interacting coefficient such that $\gamma <%
\frac{1}{2\beta_{\lim }} $. The cosmological parameters are assumed
to be $\Omega_{m_{0}}=0.315$, $%
q_{0}=-0.558$, $\Omega_{\alpha }=0.1$,
$\lambda_{m}=0$ and $\lambda_{H}=0.3$. The plot has been
done for several values of the holographic parameter that are stated at the
bottom of the figure.}
\label{Dp}
\end{figure}
\begin{figure}[t]
\begin{center}
\includegraphics[width=0.75\columnwidth]{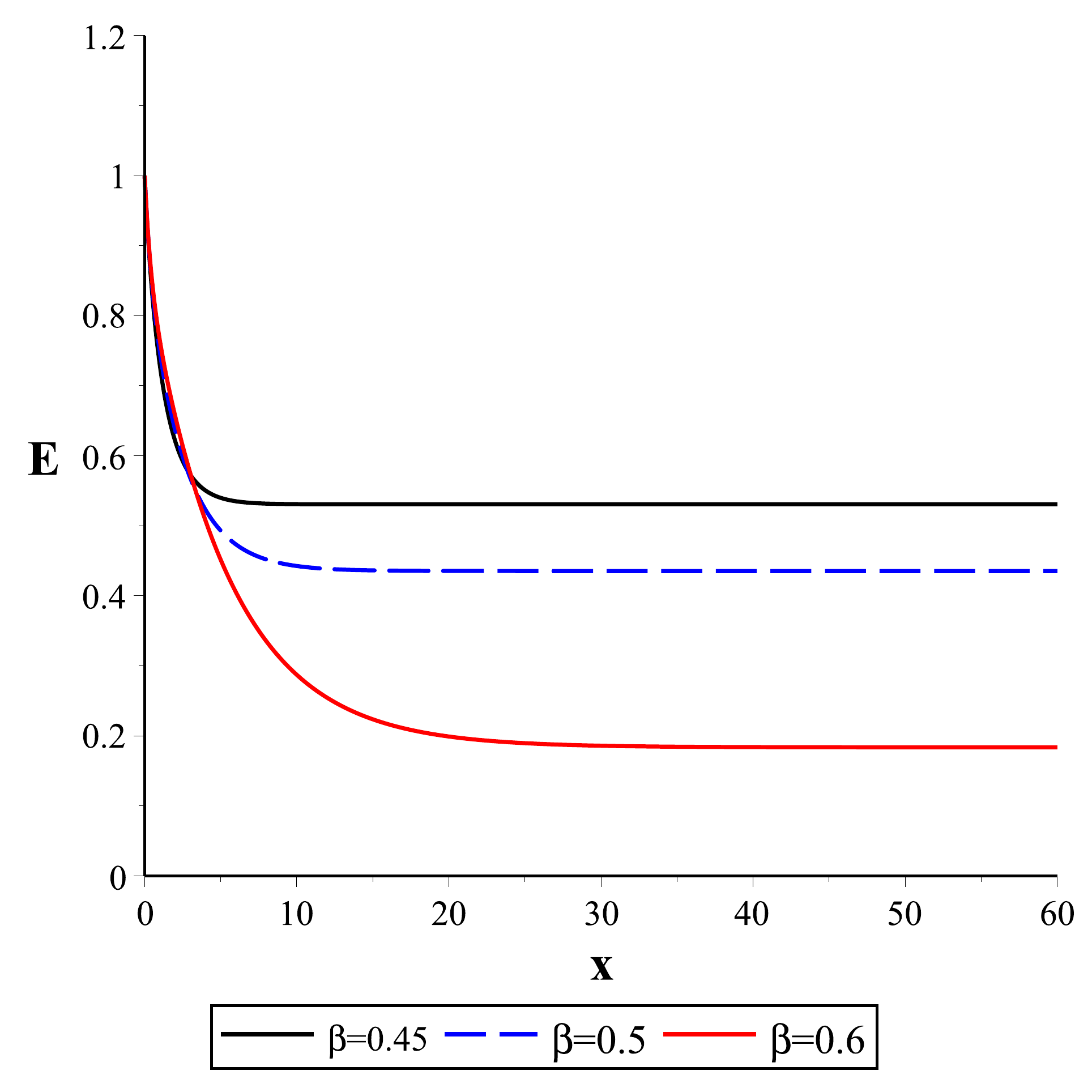}
\end{center}
\caption{Plot of the dimensionless Hubble rate $E$ against $x$, for the
positive sign of $\mathcal{D}$, and for an interacting coefficient such that $\gamma >%
\frac{1}{2\beta_{\lim }} $. The cosmological parameters are assumed
to be $\Omega_{m_{0}}=0.315$, $%
q_{0}=-0.558$, $\Omega_{\alpha }=0.1$,
$\lambda_{m}=0$ and $\lambda_{H}=0.6$. The plot has been
done for several values of the holographic parameter that are stated at the
bottom of the figure.}
\label{Dp1}
\end{figure}

\item {Positive discriminant ($0<\mathcal{D}$):}

The discriminant $\mathcal{D}$ is positive for $\frac{1}{2\beta_{\lim }}<\gamma$,
 and $\beta_{+}<\beta_{\lim }<\beta $ or for $\gamma <\frac{1}{2\beta_{\lim }}$ 
 and $\beta $ satisfying $\beta_{-}<\beta $ or $\beta_{\lim }<\beta <\beta_{+}$
(cf. Eq.~(\ref{factorD}), Fig. \ref{beta_pm}, see also Fig.~\ref{Dp}, and \ref{Dp1} which will be discussed later).

The solution of the asymptotic Friedmann equation (\ref{FriedasympGB})
reads
\begin{equation}
\left\vert \frac{4\Omega_{\alpha }\sqrt{\Omega_{r_{c}}}%
E-2\gamma \beta +1+\sqrt{\mathcal{D}}}{4\Omega_{\alpha }\sqrt{%
\Omega_{r_{c}}}E-2\gamma \beta +1-\sqrt{\mathcal{D}}}\right\vert =%
\mathcal{C}_{2}\exp \left[ -\frac{\sqrt{\mathcal{D}}}{\gamma \beta }(x-x_{2})%
\right] ,  \label{solution D>0}
\end{equation}%
where
\begin{equation}
\mathcal{C}_{2}=\left\vert \frac{4\Omega_{\alpha }\sqrt{\Omega
_{r_{c}}}E_{2}-2\gamma \beta +1+\sqrt{\mathcal{D}}}{4\Omega
_{\alpha }\sqrt{\Omega_{r_{c}}}E_{2}-2\gamma \beta +1-\sqrt{\mathcal{D}}}%
\right\vert ,
\end{equation}%
$x_{2}$ and $E_{2}$ are integration constants. We can deduce that the brane
behaves asymptotically like an expanding de Sitter universe with a positive Hubble rate
\begin{equation}
E_{+}=\frac{(2\gamma \beta -1)-\sqrt{\mathcal{D}}}{4\Omega_{\alpha }\sqrt{%
\Omega_{r_{c}}}},  \label{E_{+2}}
\end{equation}%
 for $\gamma >\frac{1}{2\beta }$. While for the
case, $\gamma \leq \frac{1}{2\beta }$ the Hubble rate $E_{+}$ is negative
and the asymptotic analysis performed is no longer valid. In fact, the brane
faces a big freeze singularity where, from Eq. (\ref{diff eq with GB}), the expansion of the brane can be
approximated by
\begin{equation}
E\sim \frac{\gamma\beta}{2\Omega_{\alpha }\sqrt{\Omega_{r_{c}}}\,(x_{\mathrm{sing_{2}%
}}-x)}.  \label{solution2 D>0}
\end{equation}%
The constant $x_{\mathrm{sing_{2}}}$ stands for the \textquotedblleft
size\textquotedblright\ of the brane at this big freeze singularity $"{\mathrm{sing_{2}}}"$. The
Hubble rate Eq. (\ref{solution2 D>0}) can be integrated over time, resulting in
the following expansion for the scale factor of the brane
\begin{equation}
a=a_{\mathrm{sing_{2}}}\exp \left[ -\left( \frac{H_{0}}{\Omega_{\alpha }%
\sqrt{\Omega_{r_{c}}}}\right) ^{\frac{1}{2}}\sqrt{t_{\mathrm{sing_{2}}}-t}%
\right].
\end{equation}%
The big freeze singularity takes place at a finite scale factor, $a_{\mathrm{%
sing_{2}}}$, and a finite cosmic time, $t_{\mathrm{sing_{2}}}$. This latter
case can be removed by an appropriate choice of $\lambda_{m}$ and $\lambda
_{H}$ through the coupling $\gamma$, which allows removing the big freeze present in the
case $\mathcal{D>}0$ by making the brane asymptotically de Sitter.\\

\item {Vanishing discriminant ($\mathcal{D}=0$):}

\begin{figure}[t]
\begin{center}
\includegraphics[width=0.75\columnwidth]{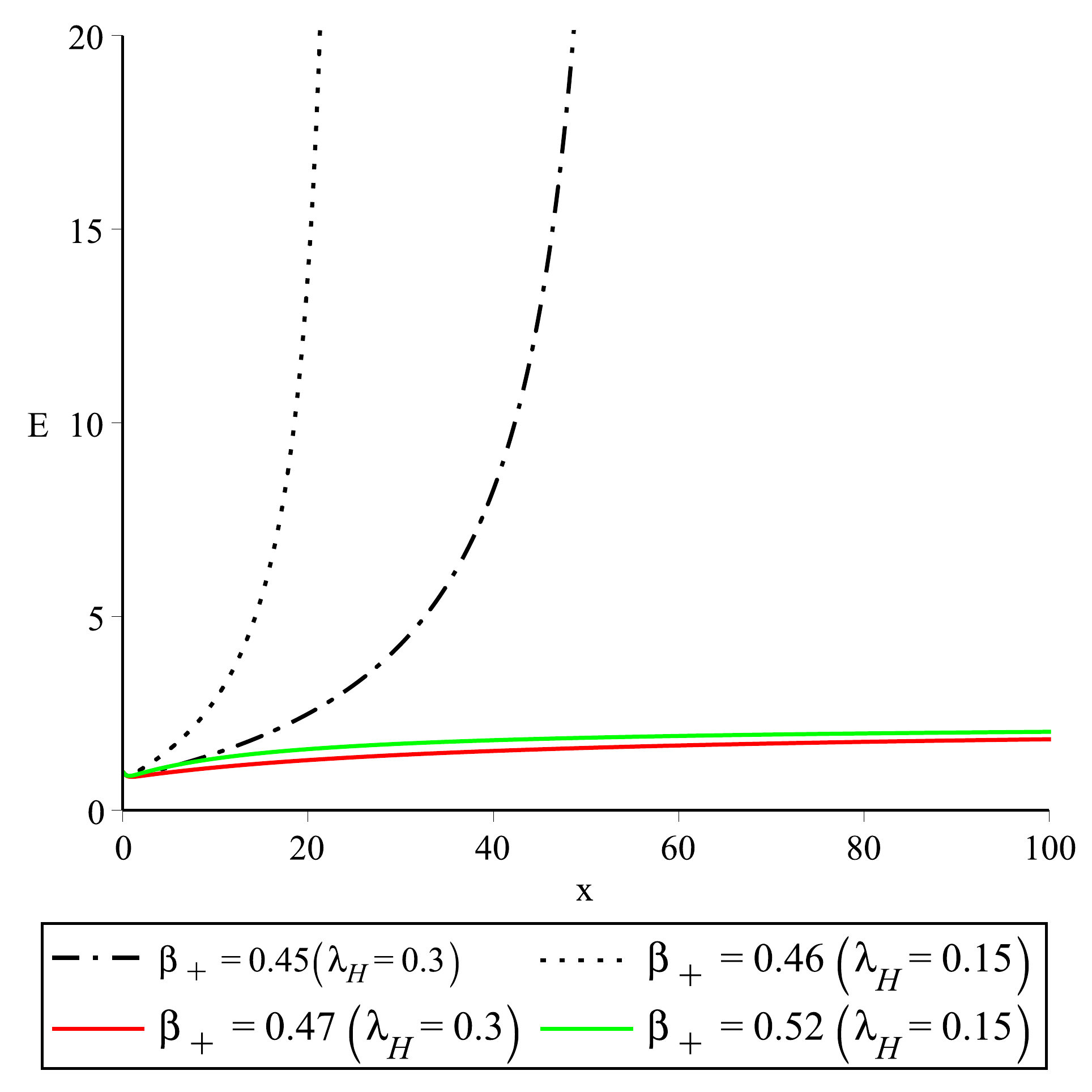}
\end{center}
\caption{{}Plot of the dimensionless Hubble rate $E$ against $x$, for
vanishing $\mathcal{D}$, and for an interacting coefficient $\gamma <\frac{1}{2%
\beta_{\lim }}$. The cosmological parameters are assumed to be $\Omega_{m_{0}}=0.315$, $%
q_{0}=-0.558$, $\Omega_{\alpha }=0.1$, and $%
\lambda_{m}=0$. The plot has been done for several values of the
holographic parameter, and the interacting parameter $\lambda_{H}$ that
are stated at the bottom of the figure.}
\label{D=0}
\end{figure}
Finally, the discriminant $\mathcal{D}$ vanishes when $\beta =\beta_{+}$ or
$\beta_{-}$. This can take place on the normal branch for $\gamma <\frac{1}{%
2\beta_{\lim }}$. The solution of the modified Friedmann equation (\ref%
{FriedasympGB}) can be expressed as
\begin{equation}
E=-\frac{1}{4\Omega_{\alpha }\sqrt{\Omega_{r_{c}}}}\left[ (2\gamma \beta
-1)-\frac{1}{\mathcal{C}_{3}+\frac{1}{2\gamma \beta }(x-x_{3})}\right] ,
\label{solution2 D=0}
\end{equation}%
where
\begin{equation}
\mathcal{C}_{3}=\frac{1}{2\gamma \beta -1+4\Omega_{\alpha }\sqrt{\Omega
_{r_{c}}}E_{3}},
\end{equation}%
$x_{3}$ and $E_{3}$ are integration constants. By taking the limit $%
x\rightarrow \infty $, the dimensionless Hubble rate (\ref{solution2 D=0})
reduces to
\begin{equation}
E_{+}=\frac{(2\gamma \beta -1)}{4\Omega_{\alpha }\sqrt{\Omega
_{r_{c}}}},  \label{E_{+3}}
\end{equation}%
$E_{+}$ coincides with the one of Eq.~(\ref{E_{+2}}) for a vanishing $%
\mathcal{D}$. The brane can be asymptotically de Sitter for the normal
branch if $\frac{1}{2\beta }<\gamma <\frac{1}{2\beta_{\lim }}$, otherwise
the Hubble rate $E_{+}$ is negative and the asymptotic analysis performed is no longer valid.
This behavior can be seen from the numerical analysis presented in Fig. {\ref{D=0}} and the brane undergoes a big freeze
singularity with an expansion given by Eq. (\ref{solution2 D>0}).

The results of the model for $Q=\lambda_{%
\mathrm{H}}H\rho_{\mathrm{H}}+\lambda_{\mathrm{m}}H\rho_{\mathrm{m}}$ are
similar to those with $Q=\lambda_{\mathrm{H}}H\rho_{\mathrm{H}}$ after making the
transformation $\lambda_{H}$ $\longrightarrow \frac{3\lambda_{H}}{%
3-\lambda_{m}}$ in the model $Q=\lambda_{\mathrm{H}}H\rho_{\mathrm{H}}$.
\begin{figure}[t]
\begin{center}
\includegraphics[width=0.75\columnwidth]{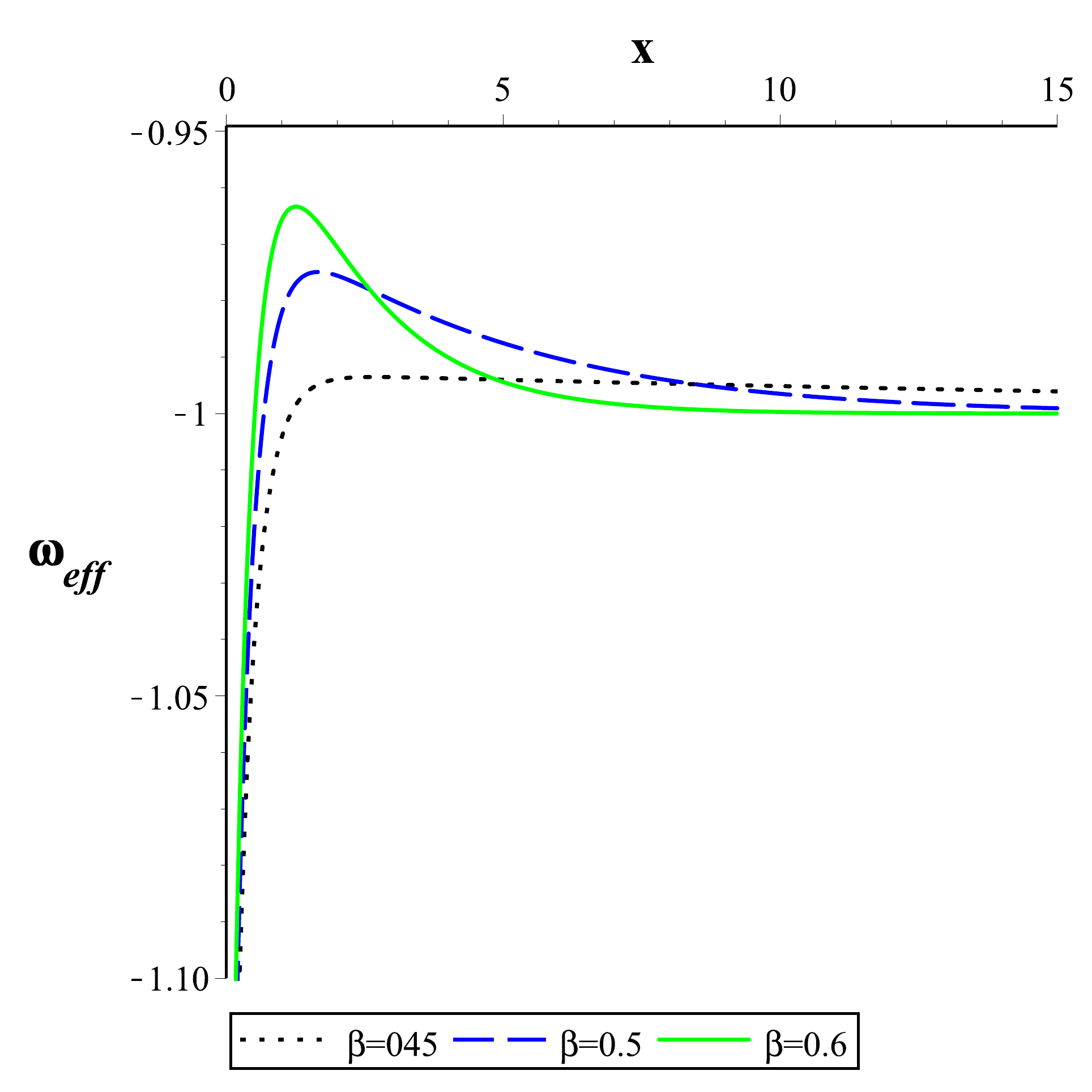}
\end{center}
\caption{Plot of the effective equation of state $\omega_{eff}$ against $x$, for $\gamma =\frac{1}{%
2\beta_{\lim }}$. The cosmological parameters are assumed to be $\Omega
_{m_{0}}=0.315$, $q_{0}=-0.558$, $\Omega_{\alpha }=0.1$. The plot has been
done for several values of the holographic parameter that are stated at the
bottom of the figure.}
\label{betalim}
\end{figure}
\begin{figure}[t]
\begin{center}
\includegraphics[width=0.75\columnwidth]{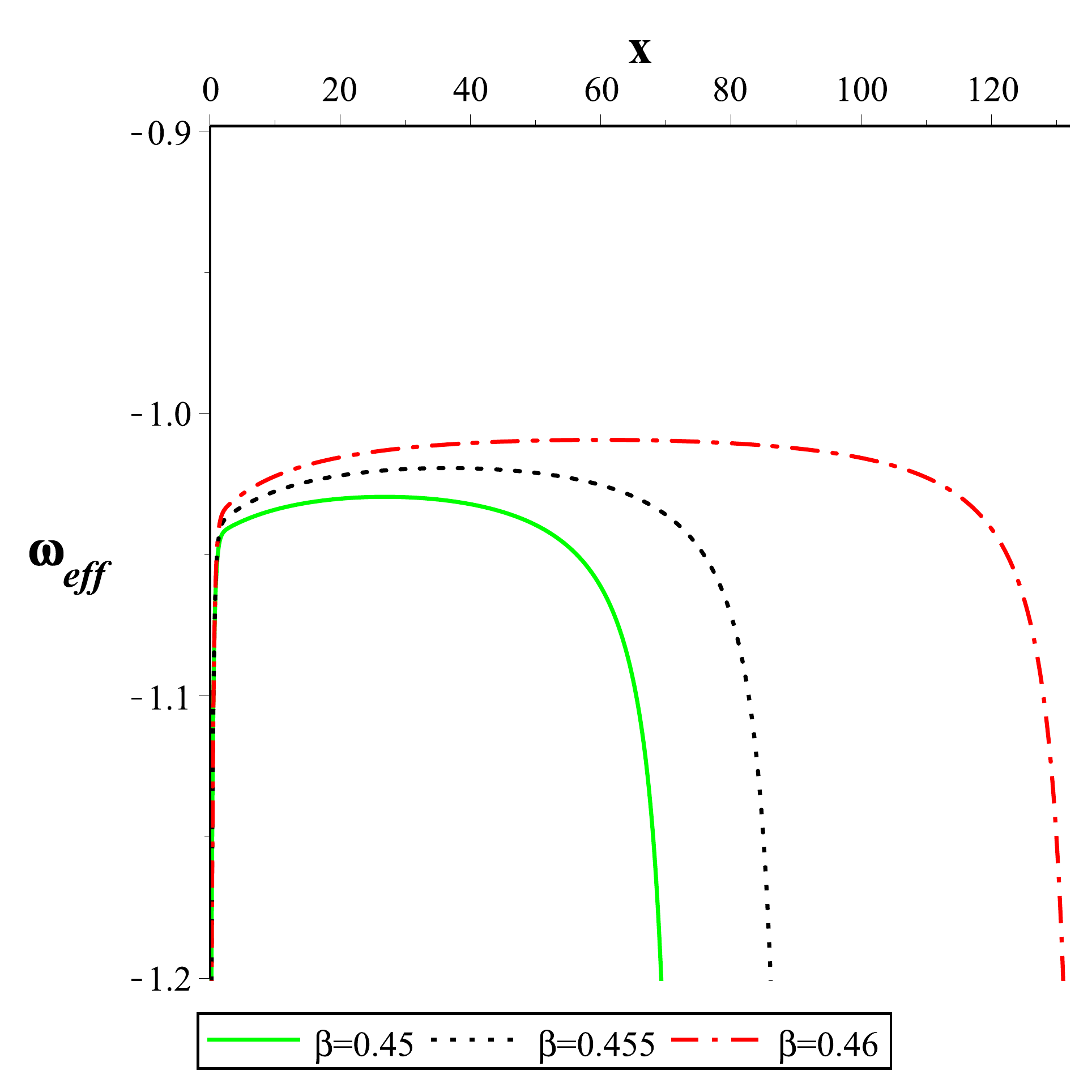}
\end{center}
\caption{Plot of the effective equation of state $\omega_{eff}$  against $x$, for the negative sign of
$\mathcal{D}$, which correspond to $\gamma <\frac{1}{2\beta_{\lim }}$. The
cosmological parameters are assumed to be $\Omega_{m_{0}}=0.315$, $q_{0}=-0.558
$, $\Omega_{\alpha }=0.1$, $\lambda_{m}=0$ and $\lambda_{H}=0.3$. The
plot has been done for several values of the holographic parameter that are
stated at the bottom of the figure.}
\label{Dnegative}
\end{figure}
\begin{figure}[t]
\begin{center}
\includegraphics[width=0.75\columnwidth]{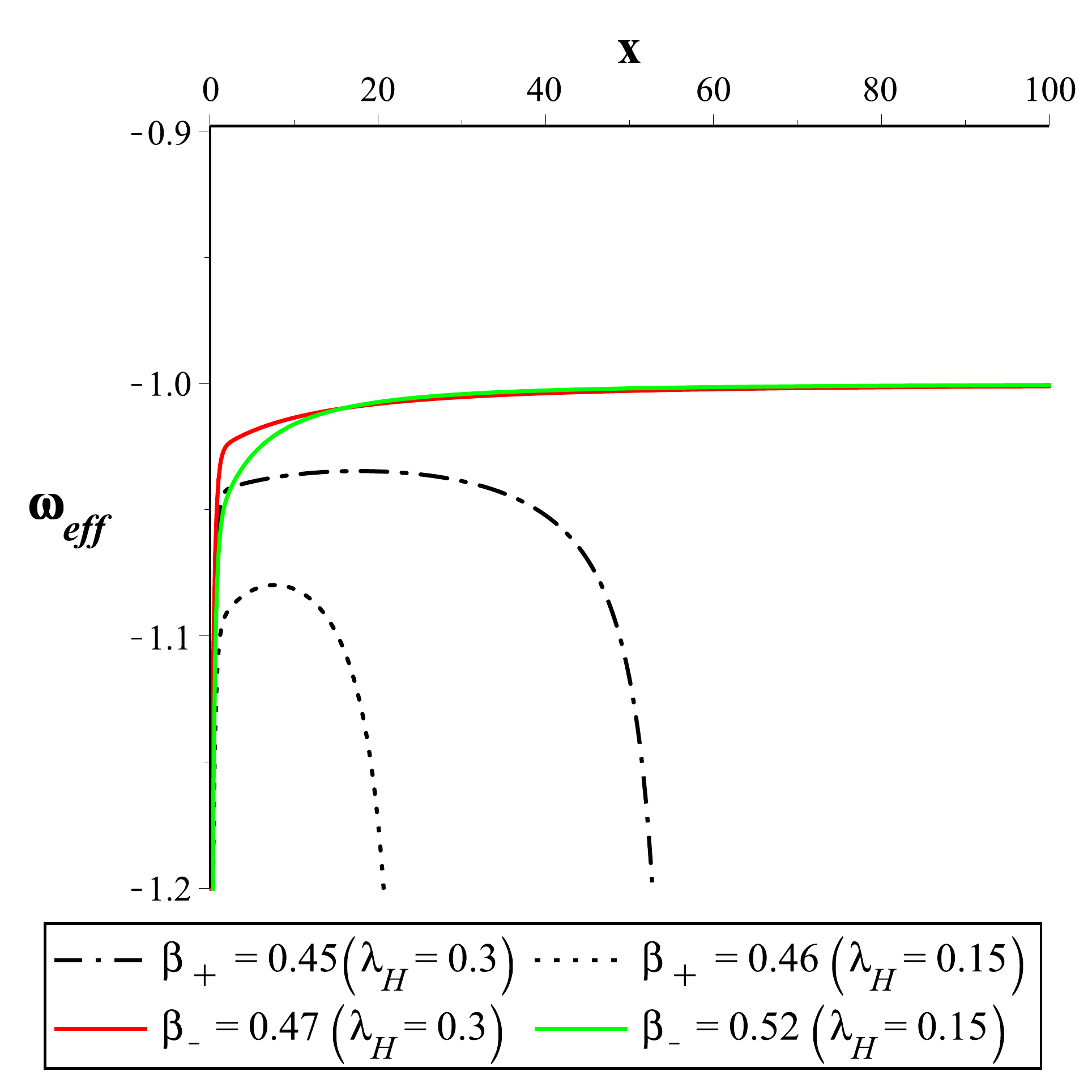}
\end{center}
\caption{Plot of the effective equation of state $\omega_{eff}$ against $x$, for the vanishing $\mathcal{D}$.
The cosmological parameters are assumed to be $\Omega_{m_{0}}=0.315$, $%
q_{0}=-0.558$, $\Omega_{\alpha }=0.1$, $\lambda_{m}=0$. The plot has been
done for several values of the holographic parameter that are stated at the
bottom of the figure.}
\label{D0}
\end{figure}
\begin{figure}[t]
\begin{center}
\includegraphics[width=0.75\columnwidth]{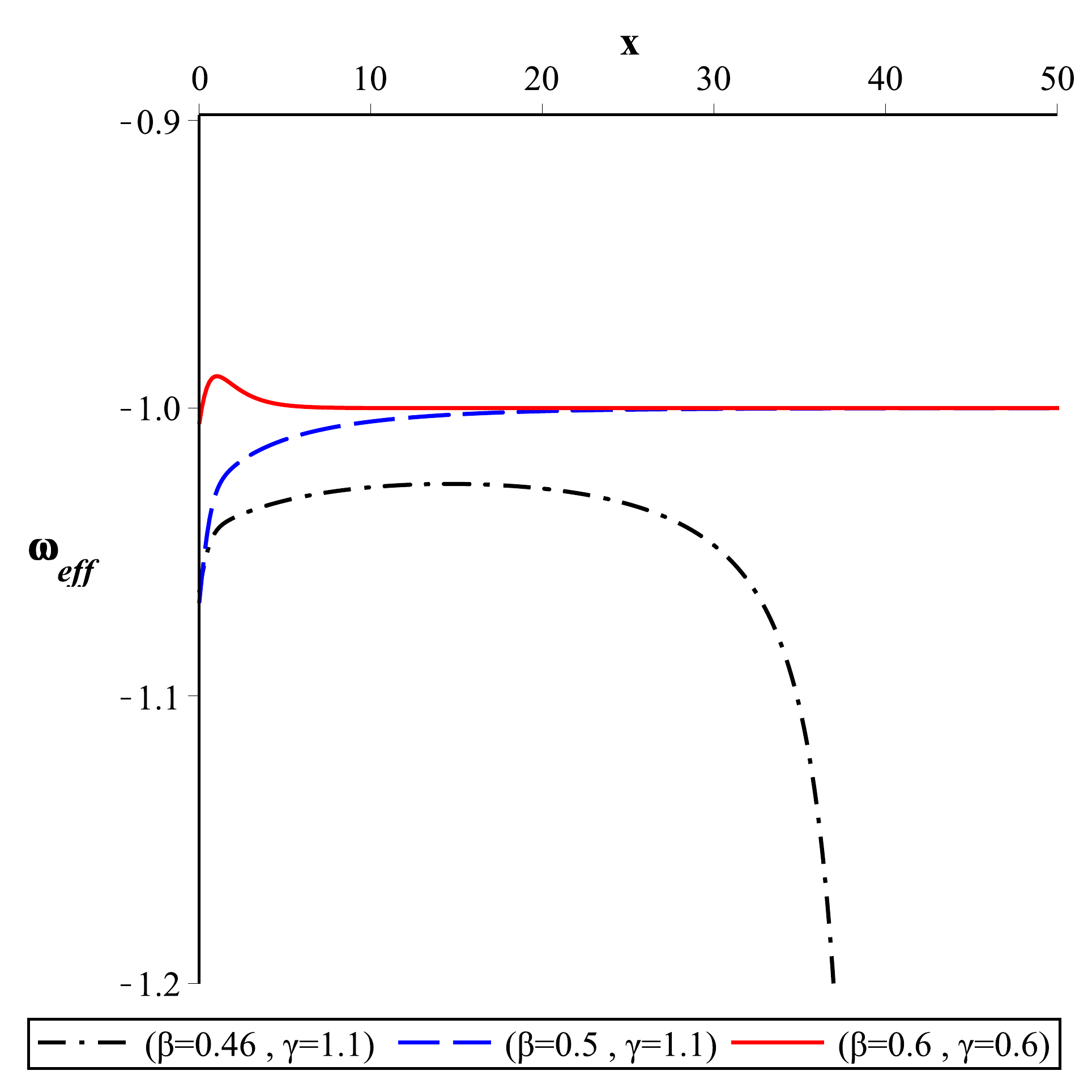}
\end{center}
\caption{Plot of the effective equation of state $\omega_{eff}$ against $x$, for the positive sign of
$\mathcal{D}$. The cosmological parameters are assumed to be $\Omega_{m_{0}}=0.315$, $%
q_{0}=-0.558$, $\Omega_{\alpha }=0.1$, $\lambda_{m}=0$. The plot has been
done for several values of the holographic parameter, and the interacting parameter $\lambda_{H}$
that are stated at the
bottom of the figure.}
\label{Dpositive}
\end{figure}

\end{enumerate}
\subsection{Numerical analysis}

The asymptotical analysis we have carried out in the previous subsection can
be completed with a numerical analysis of Eq. (\ref{variation double de E}).

Fig. {\ref{gamma0}} shows some numerical solutions for $\gamma =%
\frac{1}{2\beta_{\lim }}$ corresponding to the normal branch in which we
are interested. As we notice, the brane is asymptotically de Sitter in the
future, and the dimensionless Hubble rate $E$ approaches the value
$E_{+}=\frac{(\frac{\beta }{\beta_{\lim }}-1)-\sqrt{\mathcal{D}}}{4\Omega
_{\alpha }\sqrt{\Omega_{r_{c}}}}.$
For $\gamma \neq \frac{1}{2\beta_{\lim }}$, we discuss the numerical
analysis with respect to the sign of the discriminant $\mathcal{D}$:
(i) Fig.~\ref{Dn} shows the numerical solutions of Eq.
(\ref{variation double de E}) for a negative value of $\mathcal{D}$.
The brane expands with respect to $x$ in the future until it reaches a big
freeze singularity which is consistent with our analytical results, (ii) For a
positive sign of $\mathcal{D}$ and for $\gamma <\frac{1}{2\beta_{\lim }}$,
Fig. \ref{Dp} shows that the brane expands until it reaches a big
freeze singularity for $\beta =0.45$, and $\beta =0.52$ for $\gamma <\frac{1}{2\beta }$,
and a de Sitter solution for $\beta =1/2$ and $\beta =0.6$,
for $\gamma >\frac{1}{2\beta }$. These results are similar to
our previous analytical analysis. For $\gamma >\frac{1}{2\beta_{\lim }}$
the numerical solutions behave like a de Sitter solutions as shown in Fig. \ref{Dp1},
which is of course consistent with the analytical results,
(iii) When the discriminant $\mathcal{D}$ vanishes, the solutions are shown in
Fig {\ref{D=0}}. The brane expands with respect to $x$ in the future until
it reaches a big freeze for $\beta =0.45$, and $\beta =0.46$ for
$\gamma <\frac{1}{2\beta }$, and behaves like a de Sitter solution for $\beta =0.47$,
and $\beta =0.51$ for $\gamma >\frac{1}{2\beta }$. All of these
solutions again are consistent with our previous analytical analysis.\par
For completeness, we plot in Figs. \ref{betalim}-\ref{Dpositive} the behaviours of
the effective equation of state, $\omega_{eff}$, defined in Eq. (\ref{rhoeff1})-(\ref{rhoeff2}) where
$\rho_{\text{eff}}$ reads in this case
\begin{equation}\label{}
\rho_{\text{eff}}= 3M_p^2H_0^2\Big[\frac{1}{2}\beta\frac{dE^2}{dx}+2\beta E^2 -2\sqrt{\Omega_{\mathrm{r_c}}}(1+\Omega_{\mathrm{\alpha}}E^{2})E\Big]
\end{equation}
Fig. \ref{betalim} shows the behaviours of
the effective equation of state for $\gamma=\frac{1}{2\beta_{\textrm{lim}}}$. At the far future $\omega_{eff}$
has a phantom like even though the brane follows a de sitter behaviour.\par
Fig. \ref{Dnegative} shows the behaviours of
the effective equation of state for the negative sign of $\mathcal{D}$ i.e. $\gamma<\frac{1}{2\beta_{\textrm{lim}}}$.
The parameter $\omega_{eff}$ has a phantom like behaviour and the brane ends its expansion in a big freeze singularity.\par
Fig. \ref{D0} shows the behaviours of
the effective equation of state for the vanishing $\mathcal{D}$.
Once again the parameter $\omega_{eff}$ has a phantom like for $\frac{1}{2\beta}<\gamma<\frac{1}{2\beta_{\textrm{lim}}}$ even though the brane follows a de sitter expansion
while for $\gamma<\frac{1}{2\beta}<\frac{1}{2\beta_{\textrm{lim}}}$ the brane ends its behaviours in a big freeze singularity.\par
Fig. \ref{Dpositive} shows the behaviours of
the effective equation of state for a positive sign of $\mathcal{D}$.
The equation of state parameter $\omega_{eff}$ has a phantom like behaviour with two possible end states of the brane: (i) A de sitter behaviour of the brane for $\frac{1}{2\beta_{\textrm{lim}}}<\gamma$ e.g.
$\beta=0.6$ and $\gamma=0.6$ and for $\frac{1}{2\beta}<\gamma<\frac{1}{2\beta_{\textrm{lim}}}$  e.g. $\beta=0.5$ and $\gamma=1.1$ (ii) while for $\gamma<\frac{1}{2\beta_{\textrm{lim}}}$ the brane ends its behaviours in
a big freeze singularity.


\begin{center}
\begin{table*}[t!]
  \centering
  \begin{tabular}{ccccccc}
    \toprule
		&  &  & Without a Gauss-Bonnet term  & & & \\
		\hline\hline
 Sec. & Interacting   model  &  ~  $\lambda$~ & $\beta$   &    &Late time  behaviour   &  \\ \hline\hline
  III A &     $\lambda_{\rm m}\neq 0, \lambda_{\rm H}=0$ & {$\lambda_{\rm m}=3$} & $-$ &   &Non physical  \\

               &        & {$\lambda_{\rm m}\neq 3$} &   $\beta=\frac{1}{2}$    &   &  LR  \\
             &        & {$\lambda_{\rm m}\neq 3$} &   $\beta>\frac{1}{2}$    &    & de Sitter  \\
  &        & {$\lambda_{\rm m}\neq 3$} &   $\beta_{lim}<\beta<\frac{1}{2}$    &    & BR  \\
\hline
III B &          $\lambda_{\rm m}=0, \lambda_{\rm H}\neq 0$ &-$-$& $\beta=\beta_{LR}=\frac{3}{2(\lambda_{\rm H}+3)}$&  & LR \\
      &          & & $\beta<\beta_{LR}$& & {BR} \\
      &          & & $\beta>\beta_{LR}$& & {de Sitter} \\ \hline
III C  &    $\lambda_{\rm m}\neq 0, \lambda_{\rm H}\neq 0$ & {$\lambda_{\rm m}=3$} & $-$ &   & Minkowski  \\
               &        & {$\lambda_{\rm H}=\lambda_{\rm m}-3$} &   $-$    &   &  Non physical  \\
               &        & {$\lambda_{\rm H}\neq \lambda_{\rm m}-3$} &   $\beta=\beta_{LR}=\frac{3-\lambda_{\rm m}}{2(\lambda_{\rm H}-2(\lambda_{\rm m}+6)}$    &   &  LR  \\
               &        & {$$} &   $\beta>\beta_{LR}$    &   &  de Sitter  \\

               &        & {$$} &   $\beta<\beta_{LR}$    &   &  BR  \\

                  \hline\hline
&		& & With a Gauss-Bonnet term & & &\\
		\hline\hline
 Sec. & Interacting   model  &  ~  ${\cal{D}}$~ &$\gamma$& $\lambda$   &    Late time  behaviour   &  \\ \hline\hline
  IV A &     $\lambda_{\rm m}\neq 0, \lambda_{\rm H}=0$ & {$-$} & $1$ &$\lambda_{\rm m}=3  $& Non physical  \\

               &       &  {$-$} & $1$ &$\lambda_{\rm m}\neq3  $& Big Freeze \cite{OualiPRD85} \\

\hline
IV B $^*$ &         $\lambda_{\rm m}= 0, \lambda_{\rm H}\neq 0$ &  {${\cal{D}}<0$} & $\gamma<\frac{1}{2\beta_{lim}}$ &$- $& Big Freeze\\


               &       &  {$0\leq {\cal{D}}$} & $\frac{1}{2\beta}<\gamma\leq\frac{1}{2\beta_{lim}}$ &$- $& de Sitter \\

               &       &  {$$} & $\gamma<\frac{1}{2\beta}<\frac{1}{2\beta_{lim}}$ &$- $& Big Freeze \\

                  \hline\hline
  \end{tabular}
  \caption{Summary of the behaviours of the universe at late time, for different DM and DE interactions. ($^*$) The case $\lambda_{\rm m}\neq 0, \lambda_{\rm H}\neq 0$  is obtained by replacing $\lambda_{\rm H}$ by $\frac{3\lambda_{\rm H}}{3-\lambda_{\rm m}}$ in the expressions of $\gamma$ and ${\cal{D}}$. }
  \label{Fate-Universe}
\end{table*}
\end{center}


\section{Conclusions}

In this paper, we have presented an interacting holographic Ricci dark
energy model (IHRDE) with a cold dark matter (CDM) component in an induced gravity
brane world model. We have shown that the late time acceleration of
the universe is consistent with the current observational data with and without a Gauss-Bonnet (GB) curvature term in the
bulk action. \\

The parameter $\beta $ that characterizes HRDE is very
important in determining the asymptotic behaviour of the holographic Ricci
dark energy (HRDE) and that of the brane. The parameter $\beta$ is bounded by a limiting value, $%
\beta_{\textrm{lim}}$, that splits the self-accelerating branch from the normal one.
In this paper, we have considered only the normal branch,
i.e. $\beta >\beta_{\textrm{lim}}$, which suffers from the big rip and the
little rip singularities (see Ref. \cite{OualiPRD85}).
Assuming that, at present, our model does not deviate too much from  $%
\Lambda $CDM, the value of $\beta_{\textrm{lim}}$ is estimated to be of the order $0.44$.\\

The interaction is characterized by the quantity $Q=\lambda_HH\rho_H+%
\lambda_mH\rho_m$ where $\lambda_H$ and $\lambda_m$ are the coupling
characterizing the HRDE and the CDM interaction.\\

For a vanishing $\lambda_{H}$, the interaction model characterized only by the CDM energy does not succeed to remove
the big freeze singularity occurring in the non interacting model \cite{OualiPRD85}
with and without the GB curvature term.\\

For a vanishing $\lambda_{m}$, the model with interaction between the dark sector of
the universe can be splited into two cases:

\begin{itemize}
\item Without a GB term the IHRDE model shows that the interaction removes the
little rip singularity for $\beta =1/2$ to $\beta =%
\frac{3}{2(3+\lambda_{H})}$, and reduces the width of the big rip
singularity from $\beta_{\textrm{lim}}<\beta <1/2$ to
$\beta_{\textrm{lim}}<\beta <\frac{3}{2(3+\lambda_{H})}<1/2$. Therefore an appropriate choice of the
coupling $\lambda_{H}$ such that $\frac{3}{2(3+\lambda_{H})}<\beta
_{lim}$ avoids the big rip and the little rip and hence the IHRDE gives
a satisfactory and an alternative description of
the late time cosmic acceleration of the universe as compared to the HRDE. Indeed the later
one modifies the big rip and little rip into a big freeze one
while the former removes them definitively. Furthermore the IHRDE will have a
phantom-like behaviour even though the brane undergoes a de Sitter stage at the very late time. \\

\item With a GB term in the bulk the IHRDE model depends on the sign of the discriminant ${\mathcal{D}}$ through
the  parameter $\beta $, the GB parameter, and the coupling $\gamma $.
In the particular case $\gamma =\frac{1}{2\beta_{\textrm{lim}}}$,
the interacting model succeed in removing the big rip and little
rip singularity from the brane future evolution and it will evolve
asymptotically as a de Sitter universe  as is shown in Fig. \ref{betalim}. For $\gamma \neq \frac{1}{2\beta_{\textrm{lim}}}$, the situation depends on
the sign of the discriminant ${\mathcal{D}}$:

\begin{enumerate}
\item If ${\mathcal{D<}}0$, which correspond to $\gamma <\frac{1}{2\beta
_{\lim }}$. The brane expand in the future until it reaches a big freeze
singularity as is shown in Fig. \ref{Dnegative}.

\item If ${\mathcal{D}}=0$, two situations can be found (as is shown in Fig. \ref{D0})

\begin{description}
\item[a-] When $\frac{1}{2\beta }<\gamma <\frac{1}{2\beta_{\textrm{lim}}}$, the brane
is asymptotically de Sitter.

\item[b-] When $\gamma <\frac{1}{2\beta }<\frac{1}{2\beta_{\textrm{lim}}}$, the brane
hits a big freeze singularity.
\end{description}

\item If ${\mathcal{D}}>0$, two situations can be found (as is shown in Fig. \ref{Dpositive})

\begin{description}
\item[a-] When $\gamma >\frac{1}{2\beta_{\textrm{lim}}},$ the brane expands in the future
until it reaches a de Sitter stage.

\item[b-] When $\gamma <\frac{1}{2\beta_{\textrm{lim}}},$
the brane becomes asymptotically de Sitter in the future for $\gamma >\frac{1}{2\beta }$, while
for $\gamma <\frac{1}{2\beta }$ the brane hits a big
freeze singularity.
\end{description}
\end{enumerate}
\end{itemize}

Taking into account both couplings $\lambda_m$ and $\lambda_H$ the
description of the interaction between CDM and HRDE is
similar to the interaction characterised only by $\lambda_H$ in the HRDE by
replacing the coupling $\lambda_H$ in the interaction form $Q=\lambda_HH\rho_H$ by the
quantity $\frac{3\lambda_H}{3-\lambda_m}$. The same conclusions are obtained by requiring that the constraints
on the parameter $\gamma$ are verified for the sets of the couple $(\lambda_m,\lambda_H)$ for which $\gamma=1+\frac{\lambda_H}{3-\lambda_m}$.

\acknowledgments

The research of M. B.-L. is supported by the Basque Foundation of Science Ikerbasque. She also would like to acknowledge the partial support from the Basque government Grant No. IT956-16 (Spain) and the project FIS2017-85076-P (MINECO/AEI/FEDER, UE).

\end{document}